\documentclass{article}

\usepackage{PRIMEarxiv}

\usepackage[utf8]{inputenc} 
\usepackage[T1]{fontenc}    
\usepackage{hyperref}       
\hypersetup{
    colorlinks=true,
    linkcolor=blue,
    filecolor=cyan,      
    urlcolor=magenta,
    }
\usepackage{url}            
\usepackage{booktabs}       
\usepackage{amsfonts,amssymb,amsmath}       
\usepackage{bm} 
\usepackage{nicefrac}       
\usepackage{microtype}      
\usepackage{lipsum}
\usepackage{fancyhdr}       
\usepackage{graphicx}       
\usepackage{float}
\usepackage{wrapfig}
\usepackage{xcolor}
\usepackage{soul}
\usepackage[nameinlink]{cleveref}
\graphicspath{{Figures/}}     
\usepackage{subcaption}
\usepackage{caption}
\usepackage{vcell}
\usepackage{colortbl}
\usepackage{makecell}
\usepackage{nicematrix}

\usepackage{romannum}
\AtBeginDocument{\pagenumbering{arabic}}

\usepackage{framed,multirow}

\usepackage{pifont}
\usepackage{amsfonts,amssymb,amsmath}       
\usepackage{latexsym}
\usepackage{comment}

\usepackage{tabularray}

\UseTblrLibrary{booktabs}

\definecolor{SandyBeach}{rgb}{1,0.917,0.776}
\definecolor{NavajoWhite}{rgb}{1,0.882,0.666}
\definecolor{Tradewind}{rgb}{0.36,0.678,0.678}
\definecolor{DeYork}{rgb}{0.541,0.764,0.541}
\definecolor{ColonialWhite}{rgb}{1,0.909,0.752}
\definecolor{Grandis}{rgb}{1,0.843,0.564}
\definecolor{GoldenTainoi}{rgb}{1,0.772,0.36}
\definecolor{YellowSea}{rgb}{1,0.666,0.05}
\definecolor{WebOrange}{rgb}{1,0.647,0}
\definecolor{PinkLace}{rgb}{1,0.905,1}
\definecolor{PinkLace1}{rgb}{1,0.788,1}
\definecolor{llightgreen}{RGB}{225,250,225}
\definecolor{llightred}{RGB}{255,215,215}
\definecolor{llightcyan}{RGB}{215,255,250}
\definecolor{LavenderRose}{rgb}{1,0.607,1}
\definecolor{lgray}{rgb}{0.88,0.88,.88}
		
\definecolor{lightpurple}{RGB}{213,172,213}
\definecolor{llightpurple}{RGB}{237,221,237}
		
\definecolor{lightpink}{RGB}{255,205,213}
\definecolor{llightpink}{RGB}{255,234,237}							

\definecolor{lyellow}{RGB}{255,232,141}
\definecolor{llyellow}{RGB}{255,244,197}	

\definecolor{skyblue}{RGB}{135,206,235}
\definecolor{lskyblue}{RGB}{215,235,255}
\AtBeginDocument{\pagenumbering{arabic}}

\title{Computation-Efficient Era: A Comprehensive Survey of State Space Models in Medical Image Analysis}

\author{
  Moein Heidari\thanks{Indicates equal contribution} \\
  School of Biomedical Engineering \\
  University of British Columbia \\
  British Columbia, Canada\\
  \And
  Sina Ghorbani Kolahi\footnotemark[1] \\
  Department of Industrial and Systems Engineering \\
  Tarbiat Modares University \\
  Tehran, Iran\\
  \AND
  Sanaz Karimijafarbigloo \\
  Faculty of Informatics and Data Science \\
  University of Regensburg \\
  Regensburg, Germany\\
  \And
  Bobby Azad \\
  Electrical Engineering and Computer Science Department \\
  South Dakota State University\\
  Brookings, USA \\
  \And
  Afshin Bozorgpour \\
  Faculty of Informatics and Data Science \\
  University of Regensburg \\
  Regensburg, Germany\\
  \And
  Soheila Hatami \\
  Faculty of Mechanical Engineering \\
  Tarbiat Modares University \\
  Tehran, Iran\\
  \And
  Reza Azad \\
  Faculty of Electrical Engineering and Information Technology \\
  RWTH Aachen University \\
  Aachen, Germany\\
  \And
  Ali Diba \\
  Lunit Inc. \\
  Amsterdam, The Netherlands\\
  \And
  Ulas Bagci \\
  Department of Radiology \\
  Northwestern University \\
  Chicago, USA\\
  \And
  Dorit Merhof \\
  Faculty of Informatics and Data Science \\
  University of Regensburg \\
  Regensburg, Germany, and\\
  Fraunhofer Institute for Digital Medicine \\
  MEVIS, Bremen, Germany\\
  \And
  Ilker~Hacihaliloglu\thanks{Corresponding author: Ilker~Hacihaliloglu,  ilker.hacihaliloglu@ubc.ca}\\
  Department of Radiology, Department of Medicine\\
  University of British Columbia\\
  British Columbia, Canada\\
}

\begin{document}
\maketitle

\begin{abstract}
Sequence modeling plays a vital role across various domains, with recurrent neural networks being historically the predominant method of performing these tasks. However, the emergence of transformers has altered this paradigm due to their superior performance. Built upon these advances, Vision Transformers (ViTs) have conjoined Convolutional Neural Networks (CNNs) as two leading foundational models for learning visual representations. However, transformers are hindered by the $\mathcal{O}(N^2)$ complexity of their attention mechanisms, while CNNs lack global receptive fields and dynamic weight allocation. State Space Models (SSMs), specifically the \textit{\textbf{Mamba}} model with selection mechanisms and hardware-aware architecture, have garnered immense interest lately in sequential modeling and visual representation learning, challenging the dominance of transformers by providing infinite context lengths and offering substantial efficiency maintaining linear complexity in the input sequence. Capitalizing on the advances in computer vision, the field of medical imaging has heralded a new epoch with Mamba models. Intending to help researchers navigate the surge, this survey seeks to offer an encyclopedic review of Mamba models in medical imaging. Specifically, we start with a comprehensive theoretical review forming the basis of SSMs, including Mamba architecture and its alternatives for
sequence modeling paradigms in this context.
Next, we offer a structured classification of Mamba models in the medical field and introduce a diverse categorization scheme based on their application, imaging modalities, and targeted organs. Consequently, we explore various applications and their practical use cases in medicine, such as image segmentation, reconstruction, registration, classification, language processing, multi-modal understanding, and other medically related challenges. Finally, we summarize key challenges, discuss different future research directions of
the SSMs in the medical domain, and propose several directions to fulfill the demands of this field. In addition, we have compiled the studies discussed in this paper along with their open-source implementations on our \href{https://github.com/xmindflow/Awesome_mamba}{GitHub}\footnote{\url{https://github.com/xmindflow/Awesome_mamba}} repository. We plan to update this page regularly with the latest relevant research papers.
\end{abstract}


\section{Introduction}
\label{intro}

\begin{wrapfigure}{R}{0.42\textwidth}
	\centering
	\includegraphics[width=0.4\textwidth]{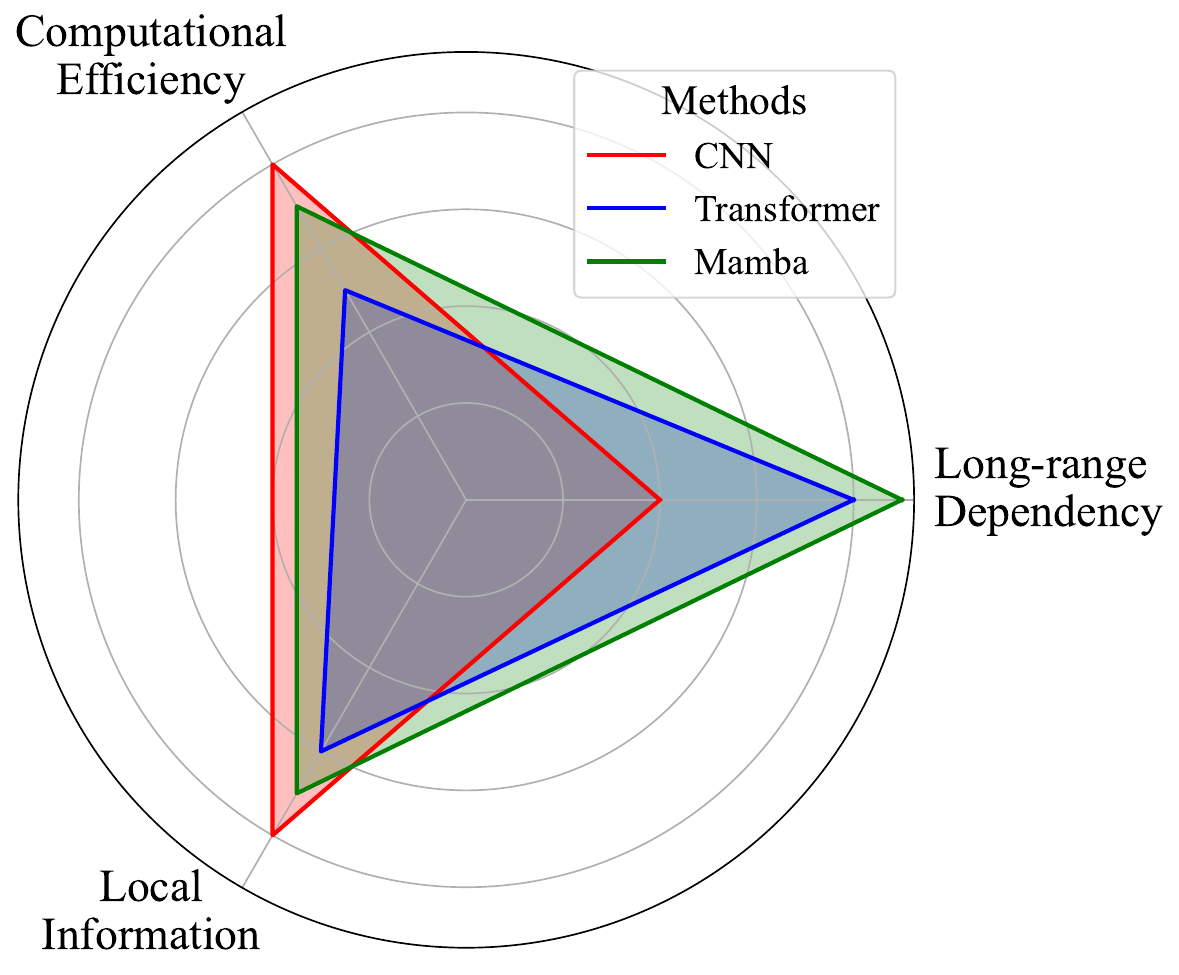}
	\caption{Algorithms and methods trilemma. Despite CNN's ability to efficiently process local information with high computational efficiency, they often struggle with capturing long-range dependencies due to their inherent architectural constraints. On the other hand, transformers excel at capturing long-range dependencies and have demonstrated high performance in various tasks. However, they are computationally expensive and potentially inefficient when processing local information. The Mamba model emerges as a promising alternative by balancing these trade-offs. It offers a method that maintains computational efficiency while effectively capturing local information and long-range dependencies.}
\label{fig:trilemma}
\end{wrapfigure}

Due to ongoing advancements in computer technology and the increasing computing capabilities of the hardware, computer-aided diagnosis (CAD) with deep learning methods has found widespread application in medical image analysis \cite{zhou2023deep,mall2023comprehensive,azad2023advances}. In the medical domain, particularly for mobile medical devices, developing artificial intelligence (AI) models with limited computational memory for deployment in resource-limited settings, such as small towns without large hospital systems or in developing countries is crucial.
Most contemporary deep learning methods, including those applied to medical imaging, are built upon the foundational architectures of convolutional neural networks (CNNs) \cite{ronneberger2015u,bhatt2021cnn,sarvamangala2022convolutional} and transformers \cite{dosovitskiy2020image,liu2023survey,khan2022transformers,chen2021transunet}. As convolutional filters designed to identify and extract necessary features from medical images, extensive research has been dedicated to CNNs, and thanks to their progressively enlarged receptive fields and excelling at obtaining local characteristics, they have been the de facto standard in CAD problems so far \cite{celard2023survey}. This research spans a range of applications, including the detection and classification of tumors \cite{kumar2021brain}, identification of skin lesions \cite{anand2023fusion}, and segmentation of target tumors \cite{balamurugan2023brain}, to list just a few examples. CNNs have also significantly enhanced the analysis of various imaging modalities in clinical medicine, including X-ray \cite{hussein2023lightweight}, CT scans \cite{jagan2023gil}, MRI \cite{ozkaraca2023multiple}, ultrasound \cite{sahu2023high}, and digital pathology \cite{meirelles2022building}. However, CNNs fail to capture long-range dependencies and spatial correlations, limiting their capacity to accurately distinguish the shape and structural details essential for effective medical image analysis across various applications \cite{li2022modeling}. On the other hand, transformers have proven to be a formidable force in the modern medical image analysis stack with tremendous impact in many applications ranging a broad area \cite{shamshad2023transformers,he2023transformers}. In contrast to CNNs, transformers, first developed for language processing and then transferred to vision domain \cite{vaswani2017attention,dosovitskiy2020image}, approach images not by processing spatial hierarchies but by treating them as sequences of patches. This criterion enhances their ability to capture global information and long-range dependencies by its elegant design of the self-attention mechanism \cite{azad2023advances}. However, the self-attention module usually introduces more computational load due to its quadratic complexity related to image size \cite{tay2022efficient} despite strategies such as Parameter-Efficient Fine-Tuning (PEFT) \cite{xin2024parameter}, and Pre-training \cite{beal2022billion} paradigm.
Despite all the progress, it is proven that both local and global information is of paramount importance in medical image processing due to the complexity of lesion distribution or organ shape \cite{heidari2023hiformer}. Furthermore, enhancing the precision of CAD frequently involves increasing the algorithmic model's parameters to boost its predictive capabilities. Yet, it is important to consider computational power and memory limitations in practical clinical and healthcare settings. For mobile health applications, it is critical to utilize models that require fewer parameters and minimal memory usage. Consequently, there is a pressing need for algorithmic models that are both efficient and perform well on future mobile medical devices~\cite{wu2024ultralight}.

Here, we highlight that these requirements present a trilemma, as current models typically make trade-offs among them. \Cref{fig:trilemma} summarizes how mainstream frameworks tackle the trilemma. Transformer models \cite{vaswani2017attention} learn the long-range feature representations well but have computational constraints and inefficiency in local information processing, CNNs, and Recurrent Neural Networks (RNNs such as Long Short-Term
Memory (LSTM) \cite{hochreiter1997long} and Gated Recurrent Unit (GRU) \cite{cho2014learning}) face limitations like solely capturing local connections, restricted memory capacity, vanishing gradients, and the incapacity to parallelize, respectively \cite{zucchet2024recurrent}. Recently, State Space-based \textbf{\textit{Mamba}} model \cite{gu2023mamba} has emerged as a promising alternative to these models to enhance computational efficiency while maintaining the ability to capture long-range dependencies and sustain high performance. They demonstrate surprisingly good results in capturing long-range and local information \cite{ma2024umamba} while scaling linearly or near-linearly with sequence length \cite{patro2024mamba}.
To date, Mamba models have proven beneficial across computer vision (CV) domains, broadly encompassing image segmentation \cite{zhu2024samba}, restoration \cite{guo2024mambair}, generation \cite{shen2024gamba}, classification \cite{chen2024rsmamba}, video \cite{li2024videomamba} and point-cloud analysis \cite{liu2024point} to name a few. The medical imaging field has recently seen a significant increase in adopting Mamba-based techniques. As illustrated in \Cref{fig:taxonomy}, extensive research focuses on applying Mamba models to various medical imaging scenarios. The recent surge of Mamba model variations has been so substantial that researchers and practitioners may struggle to keep up with the rapid pace of innovation. This trend underscores the extensive research efforts dedicated to exploring the applications of Mamba models across various medical imaging contexts. Given the current substantial attention garnered by these models, there has been a notable influx of research contributions in this domain, prompting the need for a comprehensive survey to synthesize the existing literature and offer timely insights. Hence, this survey endeavors to provide a thorough overview of the latest advancements in this category of models within medical imaging. We hope this work will point out new research directions, provide a guideline for researchers, and initiate further interest in the medical image analysis community to leverage the potential of Mamba models in the medical domain. Our major contributions are as follows:

$\bullet$~ We first present the background and theory behind
State Space Models (SSMs) briefly cover their foundation in the general formulation and the Mamba model. Then, we conduct a systematic and exhaustive examination of Mamba's uses in medical imaging, offering respective taxonomies and discussions of various aspects of these methods. In particular, we delve into over 35 papers in a structured and hierarchical fashion.

\begin{figure*}[!thb]
\centering
    \begin{subfigure}[b]{0.73\textwidth}
         \centering
         \includegraphics[width=\textwidth]{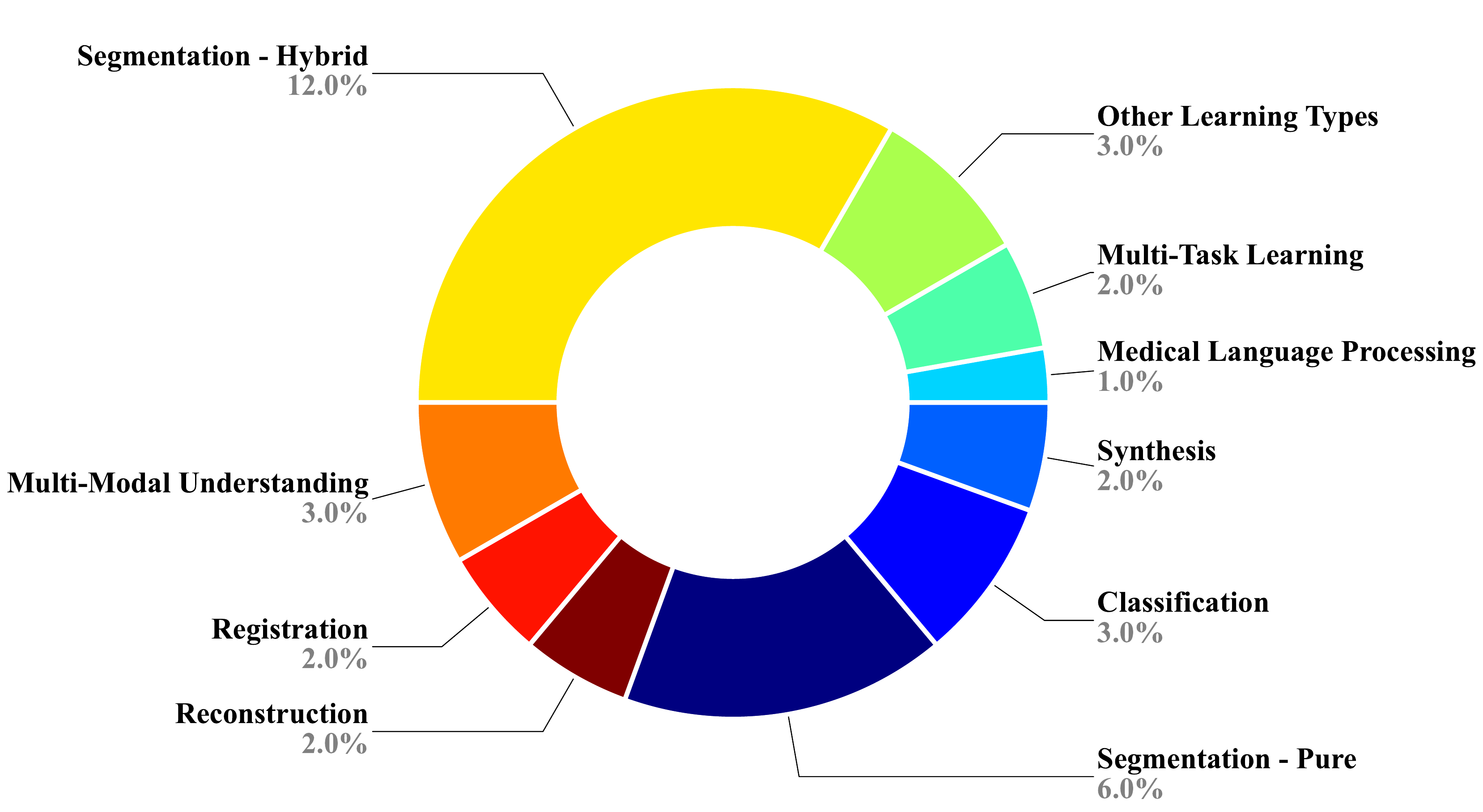}
         \caption{Applications}
     \end{subfigure}
     \hfill
     \begin{subfigure}[b]{0.59\textwidth}
         \centering
         \includegraphics[width=\textwidth]{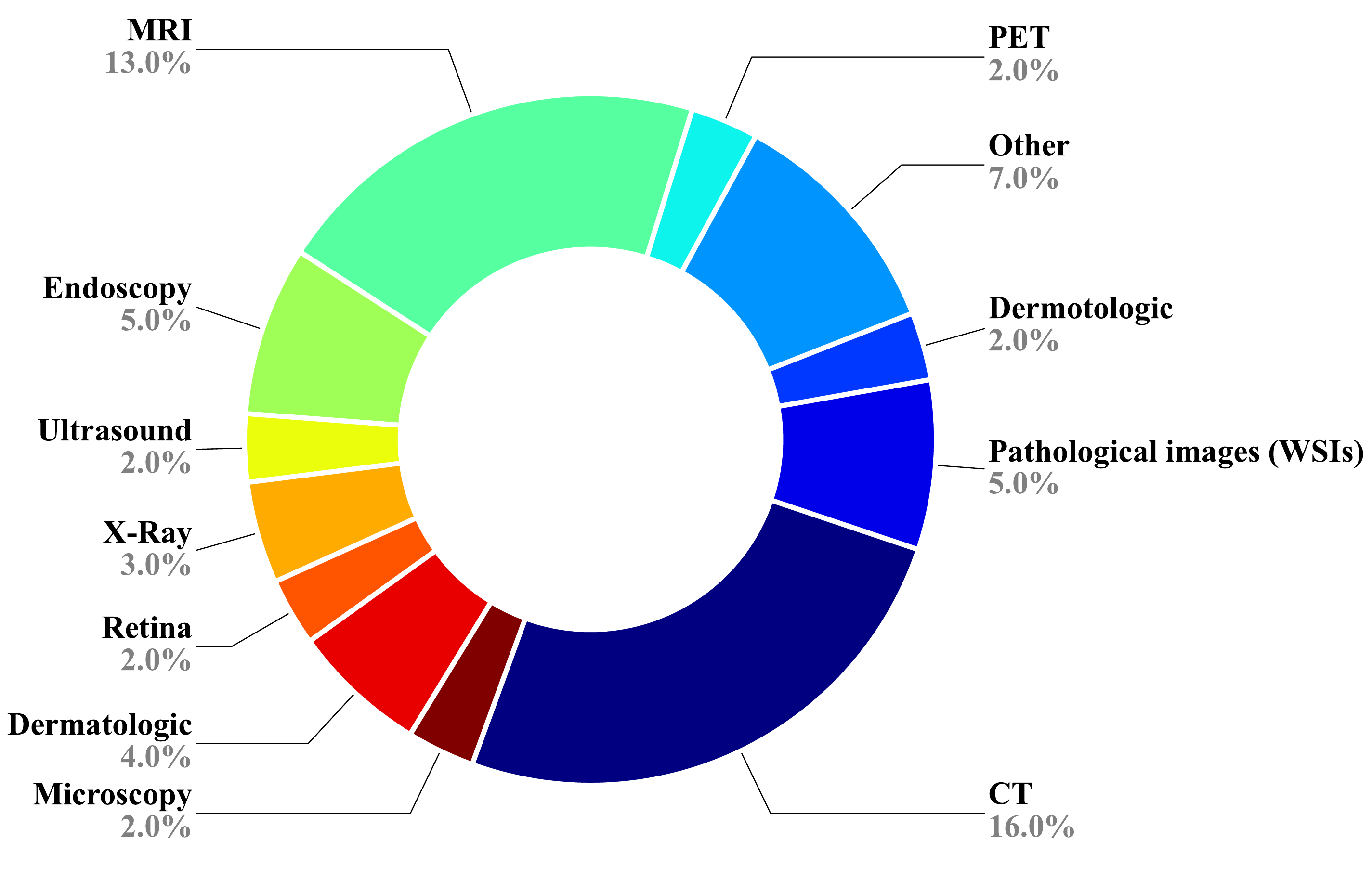}
         \caption{Modalities}
     \end{subfigure}
     \hfill
     \begin{subfigure}[b]{0.40\textwidth}
         \centering
         \includegraphics[width=\textwidth]{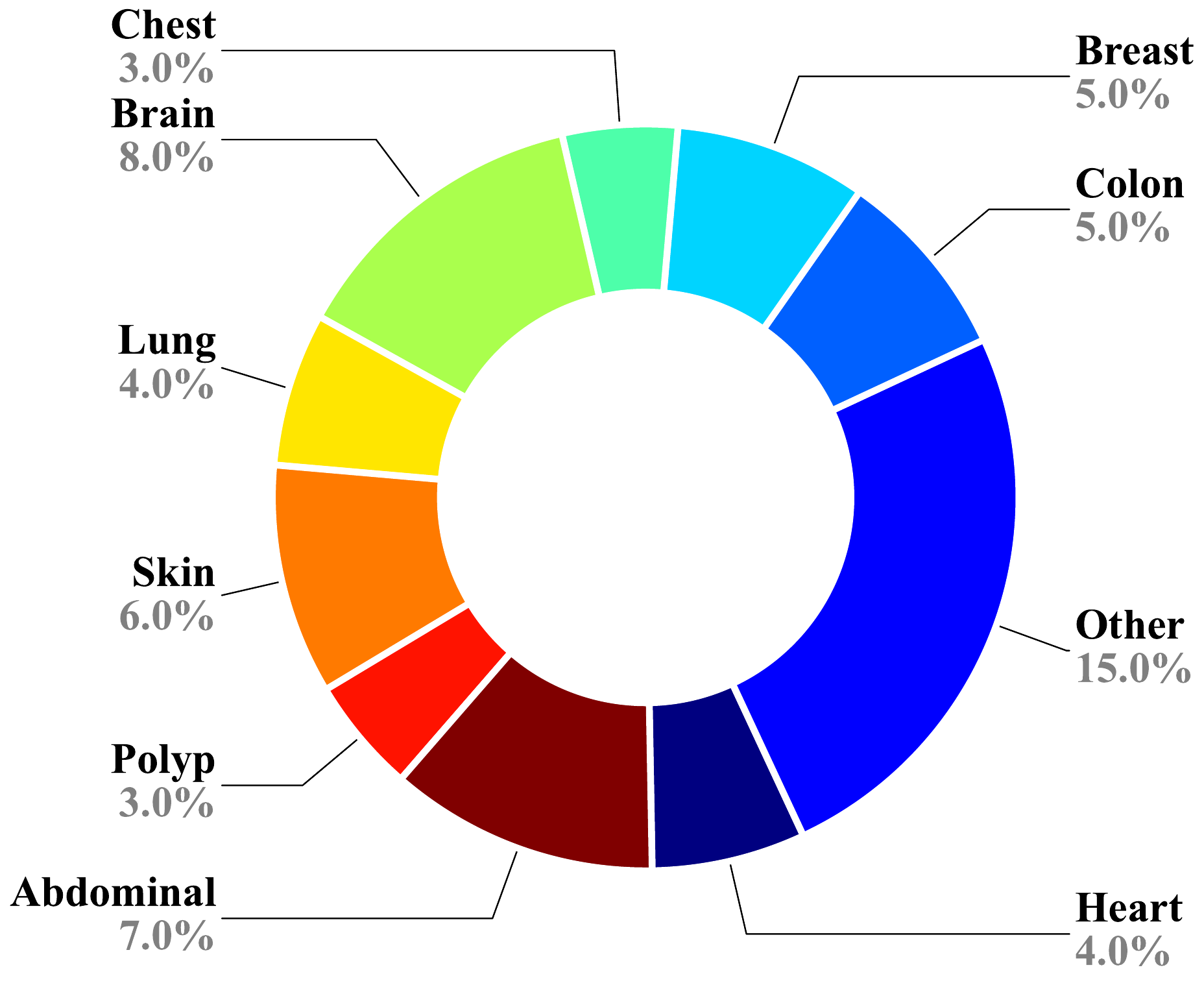}
         \caption{Organs}
     \end{subfigure}
\caption{The diagram illustrates the distribution of the analyzed papers, categorized as follows: (a) by their applications, (b) by their imaging modalities, and (c) by the type of organ studied.}
\label{fig:graphs}
\end{figure*}

$\bullet$~ Our work provides a taxonomized (\Cref{fig:taxonomy}), in-depth analysis (e.g., task/organ-specific research progress and limitations (\Cref{fig:graphs})) of these models. Specifically, we group the applications of Mamba papers into eight categories: medical image segmentation, classification, synthesis, registration, reconstruction, language processing, multi-modal understanding, and multi-task learning applications.

$\bullet$~ Finally, we address challenges and unresolved issues, highlighting emerging trends, research questions, and future directions related to the Mamba model.

\textbf{Relevant Surveys.}

Even though the Mamba network was introduced recently, related surveys and contributions to the literature have already significantly increased.
Specifically, \textit{Mamba-360} \cite{patro2024mamba} provides a holistic review of applications of SSMs across different domains while presenting the performance evaluation of SSMs compared to state-of-the-art (SOTA) transformers. Concurrently, the work of \cite{liu2024vision, zhang2024survey} carries out an extensive survey and presents a taxonomy study of Mamba models concentrating on its application across various visual tasks and data types and discusses its predecessors as well as recent advancements. Lastly, Wang et al. \cite{wang2024state} deliver a comprehensive survey of research, including both the natural language processing (NLP) and CV domains, and present statistical comparisons and analysis of these models. While some of these works touch upon the medical domain, we introduce the first comprehensive survey paper that thoroughly explores the applications of Mamba models in medical imaging. Our taxonomy systematically categorizes research on medical Mamba models and their applications, the targeted organ, and the imaging modality. We delve into SSMs' underlying concepts and theoretical principles and provide an extensive and up-to-date review of recent advancements in medical Mamba models (until June 2024), presented in \Cref{fig:taxonomy}. 

\textbf{Search Strategy.}

We performed a detailed search on DBLP, Google Scholar, and Arxiv Sanity Preserver using specialized search phrases to compile a list of academic papers. This collection encompassed peer-reviewed journals, conference and workshop papers, non-peer-reviewed articles, and preprints. Our search terms included combinations of keywords like \texttt{(mamba* $|$ deep* $|$ medical* $|$ \textbf{\{Task\}}), (mamba $|$ medical*), (state space model* $|$ medical* $|$ image* $|$ \textbf{\{Task\}}*), (V(U)mamba* $|$ vision* $|$ \textbf{\{Task\}}* $|$ medical*)}, with \textbf{\{Task\}} representing a specific application discussed in this review (refer to \Cref{fig:taxonomy}). To select the most pertinent papers, we thoroughly assessed their originality, contributions, and impact, giving preference to pioneering works in medical imaging. Based on these evaluations, we selected the most highly ranked papers for detailed analysis.

\textbf{Paper Organization.}

The structure of the remaining sections of the paper is as follows: \Cref{sec:preliminary} presents the background theory behind SSMs, covering Mamba architecture and beyond. \Cref{sec:redesign} discusses diverse variants of the Mamba model, which guide most medical domain contributions, highlighting its relevance in clinical settings. From \Cref{sec:segmentation} through \Cref{sec:multi-task}, we extensively cover the use of Mamba across various medical imaging tasks, detailing technical innovations and primary applications as illustrated in \Cref{fig:taxonomy}. In \Cref{sec:analysis}, we comprehensively discuss the approaches presented in this work and offer detailed comparisons that suggest insights into key aspects such as contributions and core ideas. Finally, we offer a detailed discussion on open challenges and potential research directions
of Mamba models in medical domain in \Cref{sec:future-direction}, and \Cref{sec:conclusion} provides a conclusion to this study.

\section{Preliminaries}
\label{sec:preliminary}

This chapter defines the necessary preliminaries of SSMs and their key features. First, a comprehensive background of SSMs is presented. Then, a unified sequence modeling formulation is introduced, describing the fundamental objectives of SSMs in processing 1-D input sequences irrespective of the input type. This is followed by an overall SSM formulation from the continuous-time system to the discretized form based on S4~\cite{gu2022efficiently}. Next, the differences between SSMs and transformers in context learning are explored, highlighting the weaknesses of SSMs in this area that led to the development of Mamba. Lastly, the architecture of Mamba is thoroughly examined.

\subsection{Background on State Space Models}
The foundational concept of the SSM is rooted in Kalman filtering~\cite{kalman1960new}, which introduces a linear filtering and prediction method divided into prediction and correction steps. The prediction phase estimates the current state based on the previous state, and the correction phase refines this estimate by assimilating the observed state, aiming for an optimal state estimation. This mathematical framework describes dynamic systems through first-order differential equations for continuous-time systems or difference equations for discrete-time systems, representing internal state evolution and output relations in matrix and vector formats, thus accommodating multivariable systems. Enhancing this model, Gu et al.~\cite{gu2021combining} developed the Linear State Space Layer (LSSL), which synergizes the strengths of recurrent neural networks, temporal convolutional networks, and neural differential equations, overcoming their individual limitations in terms of model power and computational efficiency.

Like the challenges, RNNs face, SSMs struggle with vanishing or exploding gradients, particularly when modeling extensive sequences. Addressing this, the HiPPO model~\cite{gu2020hippo} integrates concepts of Recurrent Memory with Optimal Polynomial Projections, significantly boosting the performance of recursive memory crucial for handling long sequences and dependencies. The S4 model~\cite{gu2022efficiently}, a novel sequence modeling approach, employs three innovative mechanisms to manage long dependencies effectively: the Higher-Order Polynomial Project Operator (HiPPO) for effective signal history retention, Diagonal Plus Low-Rank Parametrization for stabilizing the state transition matrix (A), and efficient kernel computation using Fast Fourier Transforms (FFTs), which optimizes computational complexity to $O(N\log(N))$.

\begin{figure}[t]
    \centering
    \includegraphics[width = \textwidth]{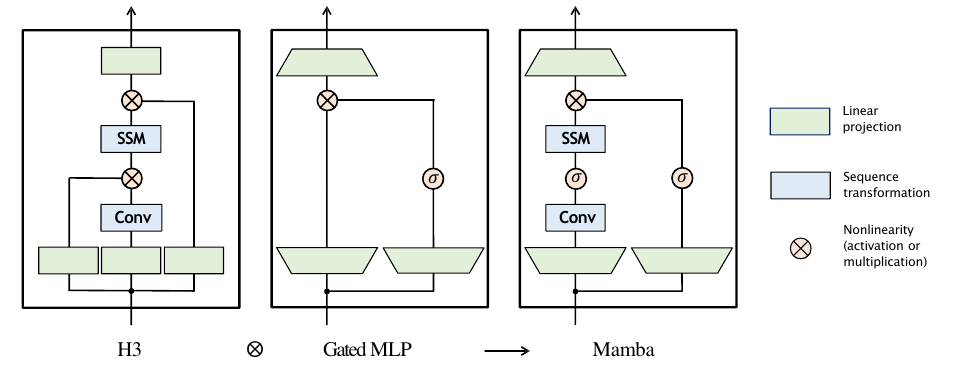}
    \caption{The Mamba block (right) is a streamlined component that combines the H3 (left) and MLP (center) blocks. From~\cite{gu2023mamba}}
    \label{fig:main-mamba}
\end{figure}

H3~\cite{fu2023hungry} identifies and addresses two significant challenges traditional SSMs encounter: difficulty retaining information from earlier tokens and comparing tokens across different sequences. To overcome these, H3, as illustrated in \autoref{fig:main-mamba}, proposes a novel approach by stacking SSMs with multiplicative interactions between input and output projections, enhanced by a FlashConv for training efficiency and a State-Passing Algorithm for scaling, which together improve memory retention and facilitate cross-sequence comparisons.

Exploring the initialization and parameterization of SSMs, Gu et al.~\cite{gu2022parameterization} and Gupta et al.~\cite{gupta2022diagonal} delve into Diagonal State Space Models (DSSM), demonstrating the critical importance of proper initialization for performance optimization. Gupta et al.~\cite{gupta2022diagonal} further present the DSS model as a simpler yet effective alternative to structured SSMs like S4, capable of achieving comparable performance without low-rank modifications.

Expanding the application scope of SSMs, the S5 model~\cite{smith2023simplified} extends the principles applied in RNN-based multiple-input multiple-output Linear Dynamical Systems to SSMs, allowing for the concurrent processing of multiple inputs and outputs. Addressing training stability, Long Convolution~\cite{fu2023simple} notes the dependence of SSMs on complex mathematical frameworks and proposes the direct parametrization of convolutional kernels to stabilize deep network training. This idea parallels innovations such as Structural Global Convolution (SGConv)~\cite{li2022makes} and Hyena Hierarchy (HH)~\cite{poli2023hyena}, which seek to address the perplexity gap observed with attention-based transformers by employing sub-quadratic convolutions and data-gating mechanisms.

Moreover, models such as RWKV~\cite{peng2023rwkv,duan2024visionrwkv}, Mega~\cite{ma2023mega}, GSS~\cite{mehta2022long}, and RetNet~\cite{sun2023retentive} illustrate the adaptive use of SSM principles. RWKV merges the benefits of transformers and RNNs, offering efficient parallelizable training and inference. RetNet, termed Retentive Network, focuses on combining training parallelism with low-cost, high-performance inference across varying computational paradigms. Gated State Spaces (GSS) further enhance this by incorporating gating units to reduce dimensionality during FFT operations, optimizing the efficiency for longer input sequences.

This enhanced detailing provides a clearer understanding of the developments in SSMs and highlights their broad applicability and ongoing evolution, as comprehensively overviewed in additional literature~\cite{cirone2024theoretical, alonso2024state}. These insights are foundational for advancing the application and theoretical understanding of SSMs in handling long sequential data.

\textbf{SSMs General Formulation}

In this subsection, we introduce the general form of causal sequence modeling tasks with one-dimensional (1-D) sequence inputs, using language modeling as an illustrative example. This framework helps establish the foundational goals of SSM applications in 1-D sequence modeling.

Sequence modeling can be viewed as a mapping between input and output signals, represented by the function \( f \):
\begin{equation}
\textbf{y}(t) = f(\textbf{x}(t), \dots, \textbf{x}(t-T); \theta),
\end{equation}
where \( \textbf{y}(t) \) is the output at time \( t \), and \( \textbf{x}(t), \dots, \textbf{x}(t-T) \) are the input signals up to time \( t \). The parameters \( \theta \) are optimized for specific tasks to enhance model performance. Due to the extensive search space of \( f(\cdot; \theta) \), various parameterizations are used to simplify the problem. The architecture of \( f(\cdot; \theta) \) significantly affects its ability to learn data structures effectively.

The model \( f(\cdot; \theta) \) learns by being exposed to input-output pairs \( (\textbf{x}(t), \textbf{y}(t)) \) for all \( t \), and iteratively updates parameters \( \theta \) to minimize the loss function \( \mathcal{L}(\cdot) \) as follows:
\begin{equation}
\bm{\min_{\theta}} \; \mathcal{L}(\textbf{y} - f(\textbf{x}; \theta)).
\end{equation}
In language modeling, inputs \( \textbf{x}(t) \) are tokenized sentences, and outputs \( \textbf{y} \) are shifted versions of the inputs, forming an auto-regressive model.

\begin{figure}[t]
    \centering
    \includegraphics[width = \textwidth]{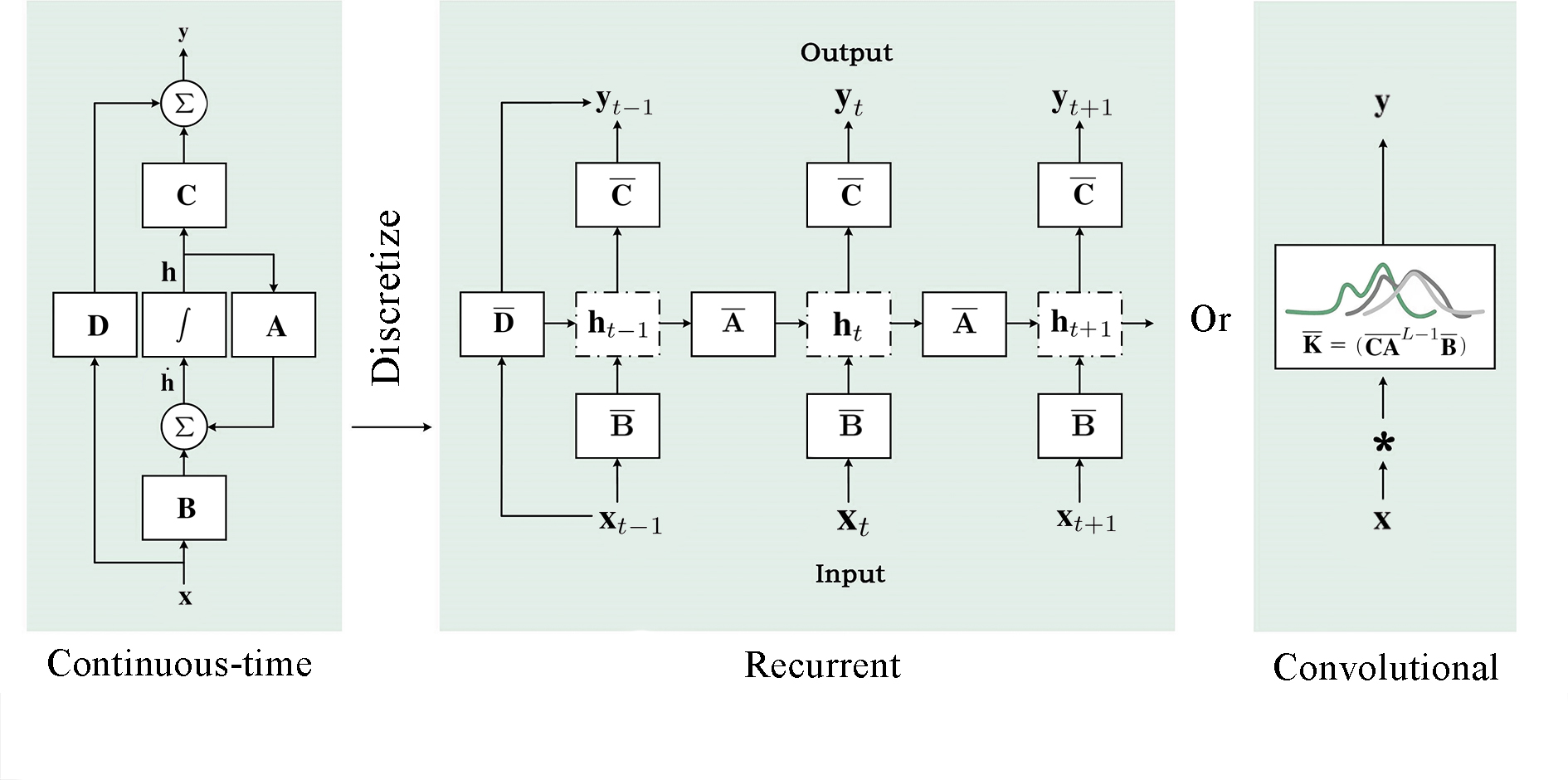}
    \caption{SSM can be represented and computed in three different forms: continuous-time, recurrent, or convolutional models. Redrawn from~\cite{gu2021combining}}
    \label{fig:ssm-rper}
\end{figure}

SSMs, inspired by the Kalman filter, represent a continuous-time system that maps a 1-D sequence from \( \textbf{x}(t) \in \mathbb{R} \) to \( \textbf{y}(t) \in \mathbb{R} \) through a hidden state \( \textbf{h}(t) \in \mathbb{R}^{N} \) as depicted in \autoref{fig:ssm-rper}. These models are typically defined by linear ordinary differential equations (ODEs):
\begin{align}
\dot{\textbf{h}}(t) &= \textbf{A} \textbf{h}(t) + \textbf{B} \textbf{x}(t), \label{eq:state}\\
\textbf{y}(t) &= \textbf{C} \textbf{h}(t) + \textbf{D} \textbf{x}(t), \label{eq:output}
\end{align}
where \( \textbf{A} \in \mathbb{R}^{N \times N}\), \( \textbf{B} \in \mathbb{R}^{N \times 1}\), \( \textbf{C} \in \mathbb{R}^{1\times N}\), and \( \textbf{D} \in \mathbb{R}^{1\times 1}\) are the state, input, output, and feed-forward matrices, respectively, assumed to be time-invariant. The state equation (\autoref{eq:state}) involves multiplying matrix \(\textbf{B}\) with the input \(\textbf{x}(t)\), followed by matrix \(\textbf{A}\) with the previous state \(\textbf{h}(t)\). The matrix \(\textbf{A}\) stores all previous information and determines the influence of the previous hidden state on the next hidden state. The matrix \(\textbf{B}\) determines how the input \(\textbf{x}(t)\) affects the hidden state. The output equation (\autoref{eq:output}) describes how the hidden state transforms into output via matrix \(\textbf{C}\) and how the input affects the output via matrix \(\textbf{D}\). In the absence of direct connections (\(\textbf{D}(t) = 0\)), the system further simplifies.

For discrete-time applications, SSMs use a timescale parameter \( \mathbf{\Delta} \) to convert continuous parameters into discrete ones. The zero-order hold (ZOH) technique is commonly used for this transformation:
\begin{equation}
\begin{aligned}
\mathbf{\overline{A}} &= \exp{(\mathbf{\Delta}\mathbf{A})}, \\
\mathbf{\overline{B}} &= (\mathbf{\Delta} \mathbf{A})^{-1}(\exp{(\mathbf{\Delta} \mathbf{A})} - \mathbf{I}) \cdot \mathbf{\Delta} \mathbf{B}.
\end{aligned}
\end{equation}
The discrete model then operates stepwise:
\begin{align}
\textbf{h}_t &= \overline{\textbf{A}} \textbf{h}_{t-1} + \overline{\textbf{B}} \textbf{x}_t, \\
\textbf{y}_t &= \textbf{C} \textbf{h}_t.
\end{align}
However, this recurrence form (\Cref{fig:ssm-rper}) does not allow for parallel computation. To address this, the model can be expanded to show each output as a function of all previous inputs:
\begin{equation}
\begin{aligned}
\textbf{y}_0 &= \textbf{C} \overline{\textbf{A}}^0 \overline{\textbf{B}} \textbf{x}_0, \\
\textbf{y}_1 &= \textbf{C} \overline{\textbf{A}}^1 \overline{\textbf{B}} \textbf{x}_0 + \textbf{C} \overline{\textbf{A}}^0 \overline{\textbf{B}} \textbf{x}_1, \\
\textbf{y}_2 &= \textbf{C} \overline{\textbf{A}}^2 \overline{\textbf{B}} \textbf{x}_0 + \textbf{C} \overline{\textbf{A}}^1 \overline{\textbf{B}} \textbf{x}_1 + \textbf{C} \overline{\textbf{A}}^0 \overline{\textbf{B}} \textbf{x}_2.
\end{aligned}
\end{equation}
Alternatively, using a convolutional approach simplifies the computation to a single convolution operation:
\begin{equation}
\begin{aligned}
\overline{\textbf{K}} &= (\textbf{C} \overline{\textbf{B}}, \textbf{C} \overline{\textbf{A}} \overline{\textbf{B}}, \ldots, \textbf{C} \overline{\textbf{A}}^{L-1} \overline{\textbf{B}}), \\
\textbf{y} &= \textbf{x} * \overline{\textbf{K}},
\end{aligned}
\end{equation}
where \( \overline{\textbf{K}} \in \mathbb{R}^{L} \) is a structured convolutional kernel that encapsulates the combined effect of the system's dynamics over the input sequence of length \( L \), and \( * \) denotes the convolution operation (\Cref{fig:ssm-rper}).

While facilitating certain computations, these models' linear time-invariant (LTI) nature introduces significant limitations in dynamic learning contexts, such as those required by deep learning frameworks. Subsequent sections of this paper will address this fundamental challenge in applying SSMs to deep learning by introducing the Mamba model. This advanced model seeks to mitigate the contextual learning deficiencies observed in traditional SSM frameworks. While the Transformer architecture stores context information in the similarity matrix, SSMs lack a similar module, adversely affecting their performance in contextual learning scenarios.

\textbf{Mamba}

To address the limitations of traditional SSMs characterized by LTI parameters, Gu et al.~\cite{gu2023mamba} have developed the Mamba architecture. This model introduces a Selective Scan Mechanism, enabling dynamic filtering of inputs based on their relevance to the task. Unlike traditional SSMs, in Mamba, the matrices \(\textbf{B}\), \(\textbf{C}\), and \(\Delta\) in S4 are functions of the inputs \( \textbf{x}(t) \), enhancing the model's responsiveness and allowing it to adapt its behavior based on the input with batch size \( B \), length \( L \), and \( D \) channels. The discretization process after incorporating the selection mechanism is as follows:
\begin{equation}
\begin{aligned}
{\overline{\textbf{B}}} &= s_B(\textbf{x}(t)),\\
{\overline{\textbf{C}}} &= s_C(\textbf{x}(t)),\\
{\Delta} &= \tau_A(\text{Parameter} + s_A(\textbf{x}(t)))
\end{aligned}
\end{equation}

where \(\overline{\textbf{B}} \in \mathbb{R}^{B \times L \times N}\), \(\overline{\textbf{C}} \in \mathbb{R}^{B \times L \times N}\), and \(\Delta \in \mathbb{R}^{B \times L \times D}\). \( s_B(\textbf{x} \) and \( s_C(\textbf{x}(t)) \) are linear functions that project the input \( \textbf{x} \) into an \( N \)-dimensional space, while \( s_A(x) \) projects the hidden state dimension \( D \) linearly into the desired dimension, connected to the RNN gating mechanism. Through these computations, the parameters \(\Delta\), \(\textbf{B}\), \(\textbf{C}\) become input functions with length \( L \), transforming the time-invariant model into a time-varying model, thus achieving selectivity.

The parameter \(\textbf{A}\) does not directly depend on the input data, but after the discretization process of the SSM, its relevance to the input can be established through the data dependency of \(\Delta\). Given that \(\textbf{A}\) is of dimension \(N\), it plays distinct roles in each dimension of the SSM. This differentiation enables precise generalization of all preceding information rather than merely compressing it.

The Mamba architecture further includes a Hardware-aware Algorithm designed to optimize computational processes by utilizing parallel scanning, kernel fusion, and strategic recalculation. These features are part of an expanded architecture that includes model dimension expansion by a factor \( E \), normalization, and residual connections. Activation functions such as SiLU or Swish improve the gating mechanisms in MLP configurations, with optional Layer Normalization to stabilize training (\Cref{fig:main-mamba}).

This innovative framework significantly enhances the adaptability and computational efficiency of SSMs, making them suitable for complex dynamic modeling tasks such as language modeling, CV, and medical image analysis.

\section{Architecture Redesign}
\label{sec:redesign}

\textbf{Scope of this section}

\begin{figure}[htbp]
    \centering
    \begin{subfigure}[b]{0.25\textwidth}
        \centering
        \includegraphics[width=0.8\textwidth]{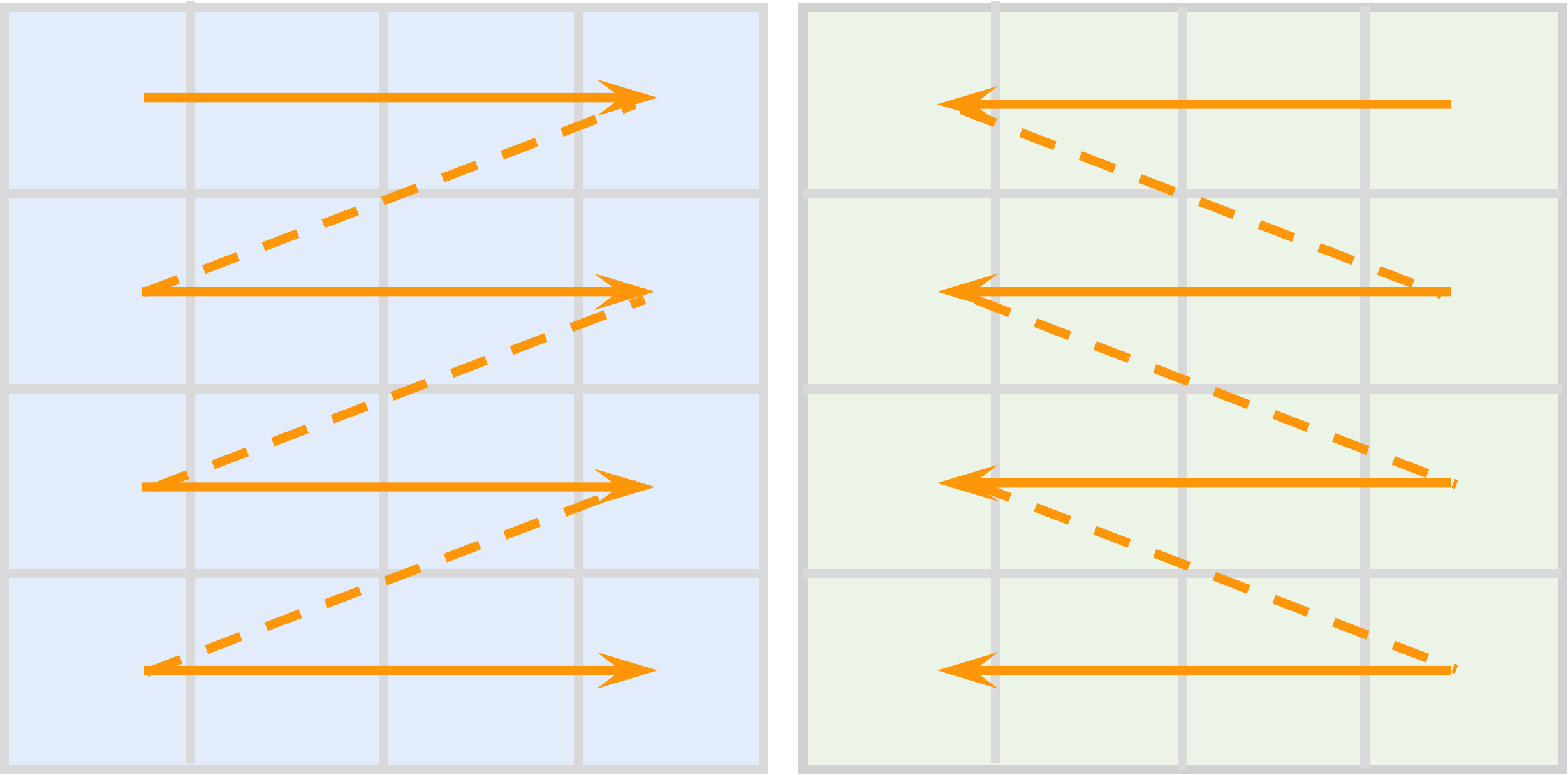}
        \caption{Bidrectional Scan~\cite{zhu2024vision}}
        \label{fig:bidrectional}
    \end{subfigure}
    \hspace{0.2\textwidth}
    \begin{subfigure}[b]{0.25\textwidth}
        \centering
        \includegraphics[width=0.75\textwidth]{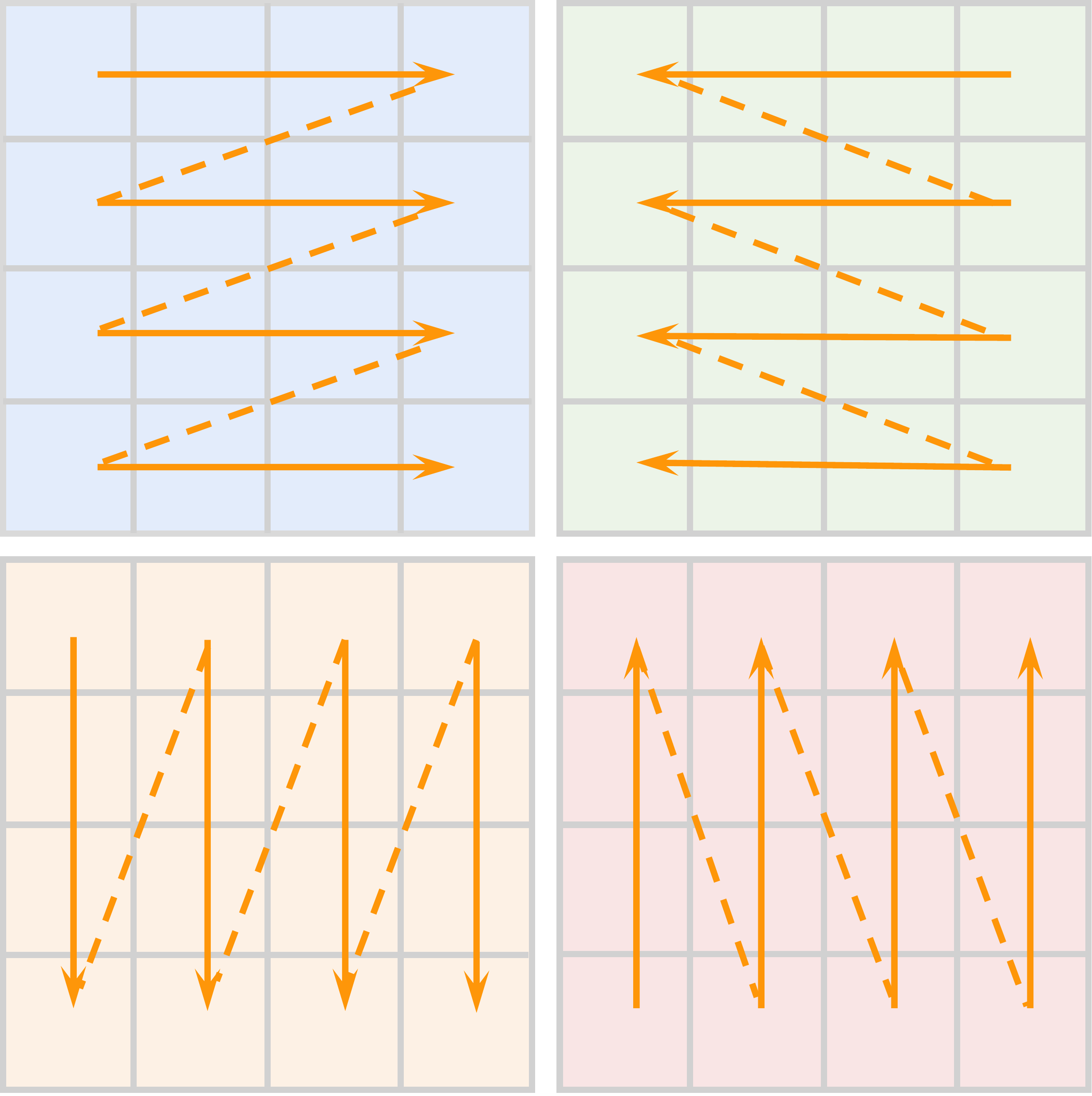}
        \caption{Cross-Scan~\cite{liu2024vmamba}}
        \label{fig:cross-scan}
    \end{subfigure}
    \hspace{0.2\textwidth}
    \begin{subfigure}[b]{0.25\textwidth}
        \centering
        \includegraphics[width=0.75\textwidth]{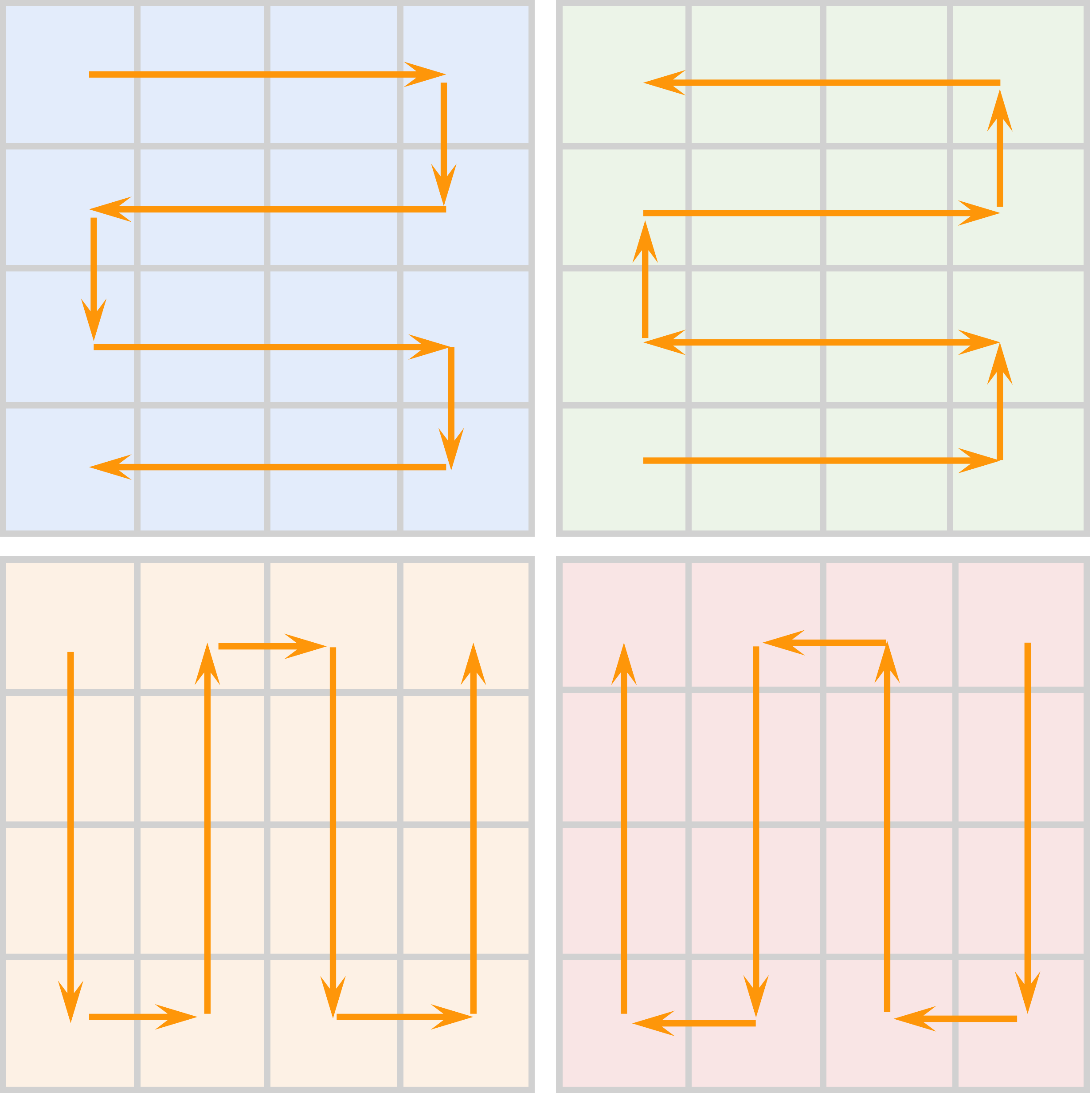}
        \caption{Continuous 2D Scan~\cite{yang2024plainmamba}}
        \label{fig:cont2d}
    \end{subfigure}
    \hspace{0.2\textwidth}
    \begin{subfigure}[b]{0.25\textwidth}
        \centering
        \includegraphics[width=0.75\textwidth]{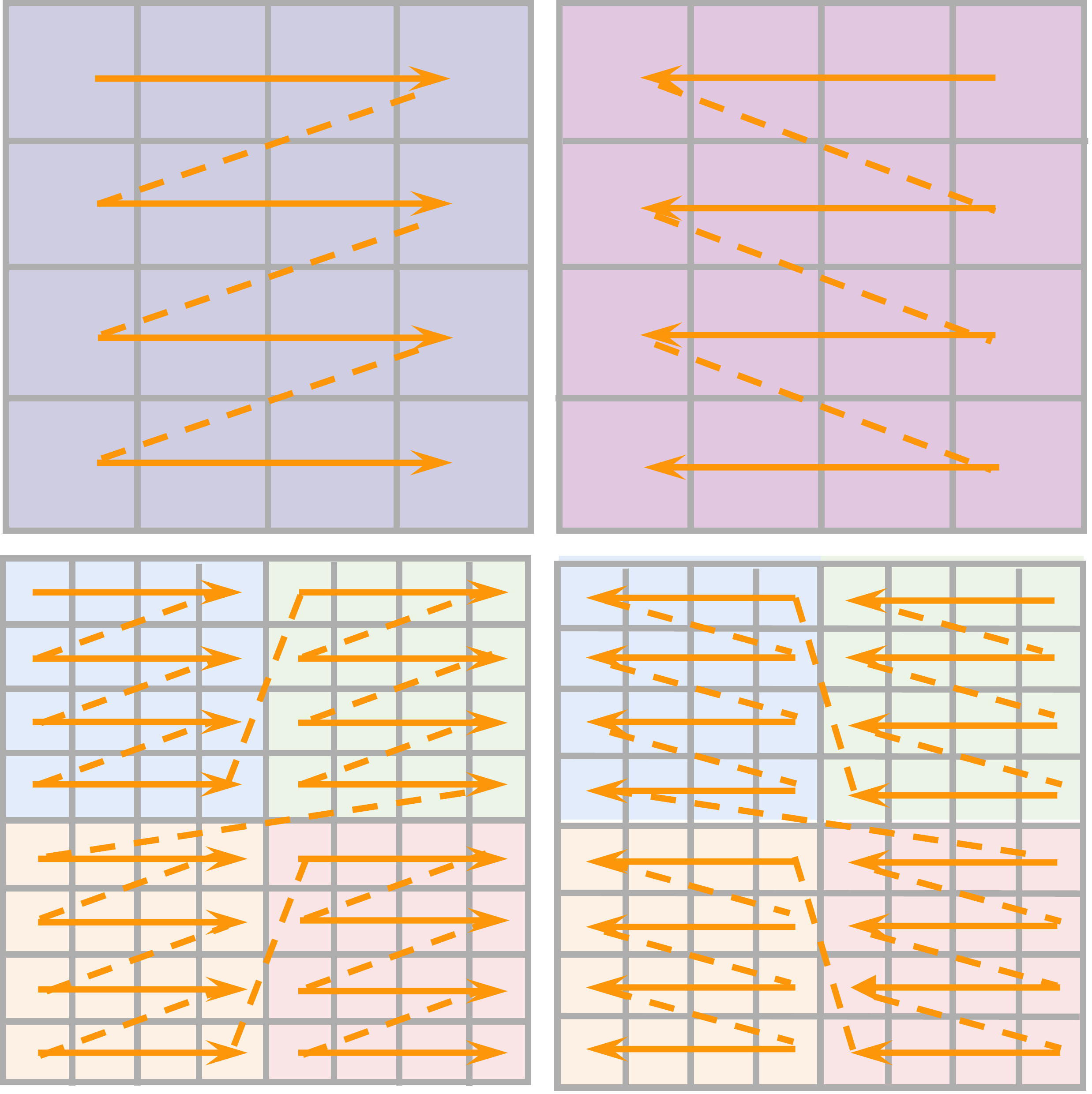}
        \caption{Local Scan~\cite{huang2024localmamba}}
        \label{fig:local-scan}
    \end{subfigure}
    
    \vskip\baselineskip 
    
    \begin{subfigure}[b]{0.45\textwidth}
        \centering
        \includegraphics[width=0.75\textwidth]{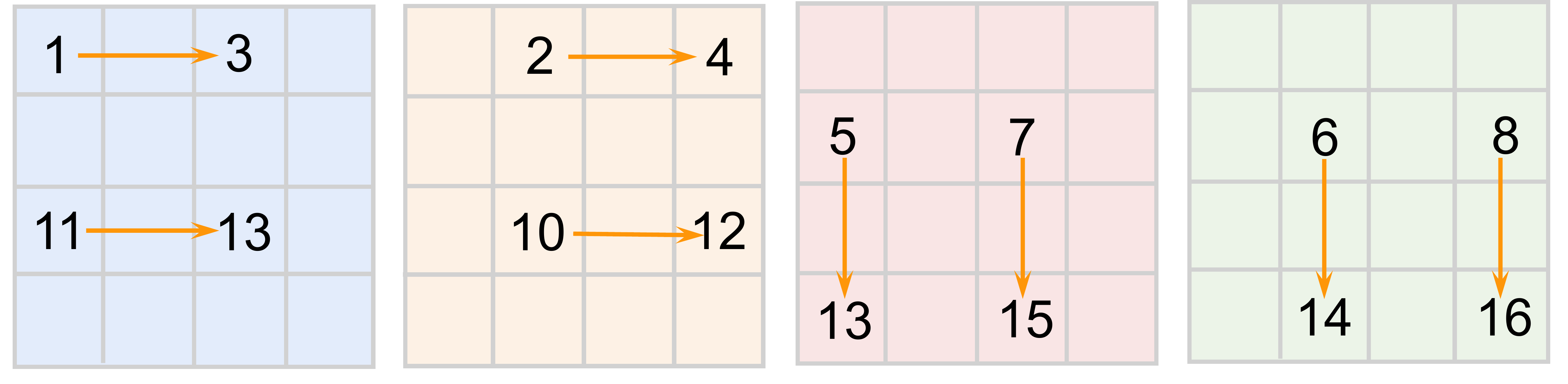}
        \caption{Efficient 2D Scanning~\cite{pei2024efficientvmamba}}
        \label{fig:ef2d}
    \end{subfigure}
    \hfill
    \begin{subfigure}[b]{0.45\textwidth}
        \centering
        \includegraphics[width=0.65\textwidth]{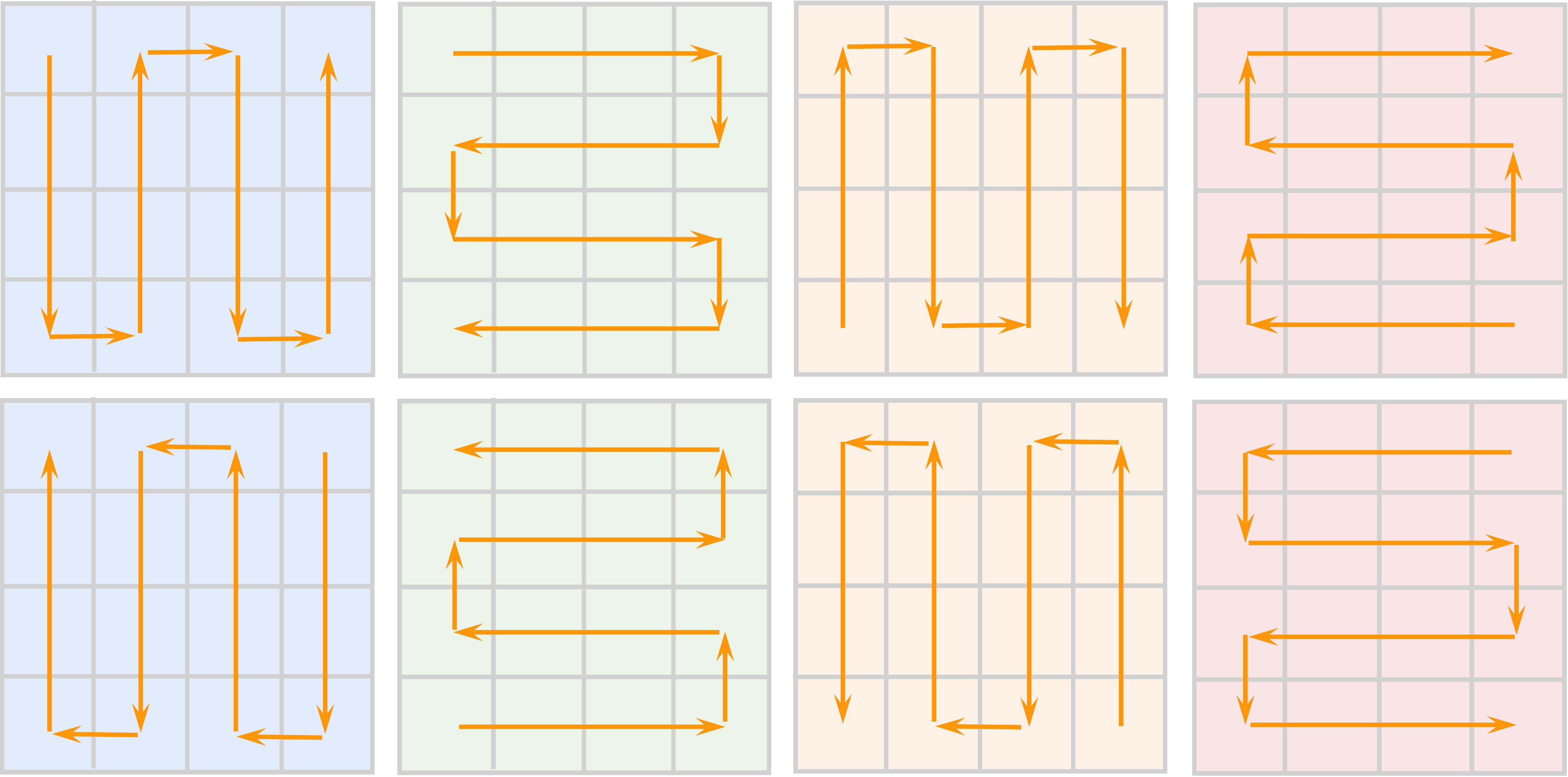}
        \caption{Zigzag Scan~\cite{hu2024zigma}}
        \label{fig:zigzag}
    \end{subfigure}
    
    \vskip\baselineskip 
    
    \begin{subfigure}[b]{0.45\textwidth}
        \centering
        \includegraphics[width=0.65\textwidth]{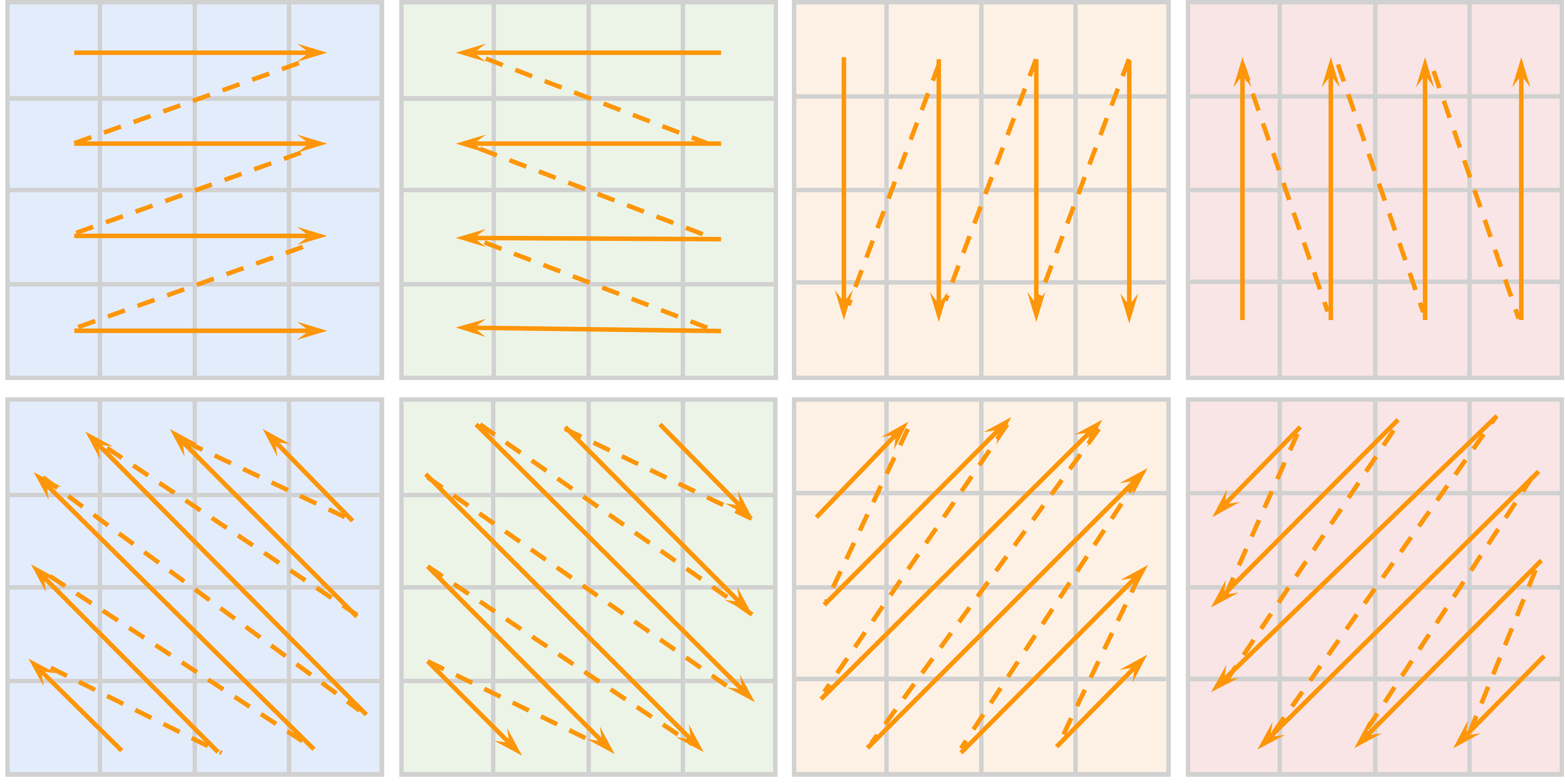}
        \caption{Omnidirectional Selective Scan~\cite{shi2024vmambair}}
        \label{fig:omnidirec}
    \end{subfigure}
    \hfill
    \begin{subfigure}[b]{0.45\textwidth}
        \centering
        \includegraphics[width=0.75\textwidth]{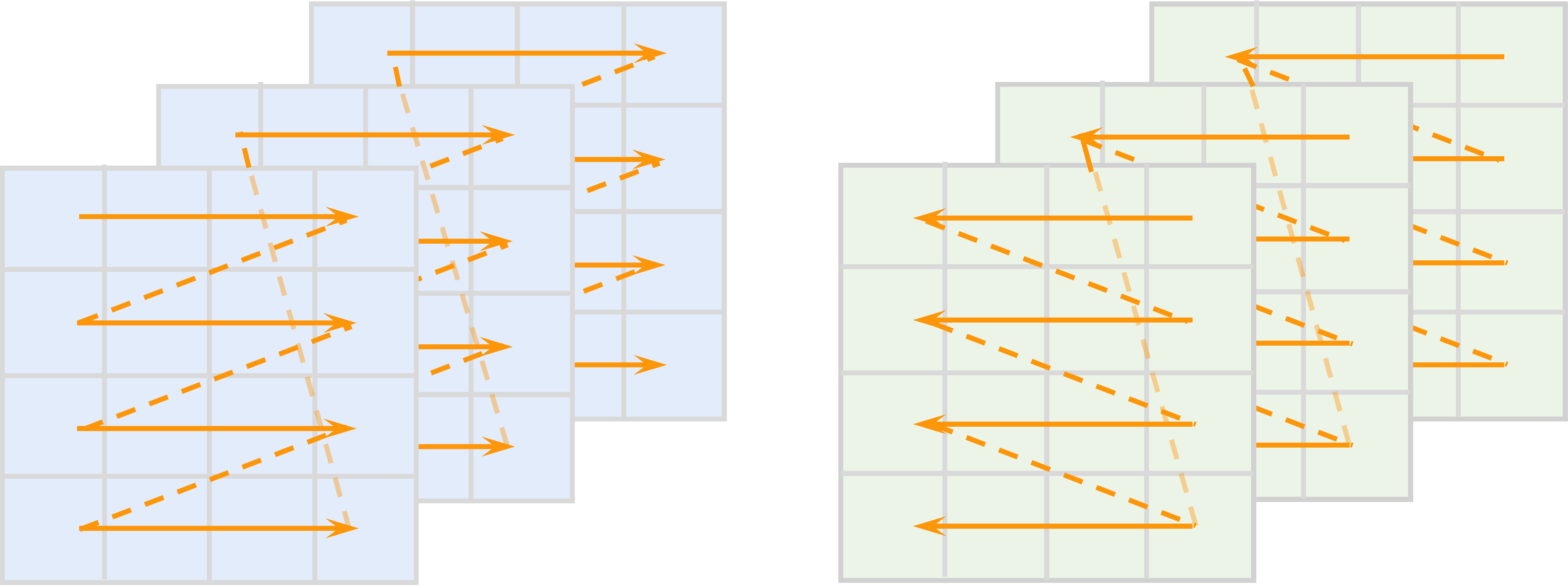}
        \caption{3D BiDirectional Scan~\cite{li2024videomamba}}
        \label{fig:3dbidirec}
    \end{subfigure}

    \vskip\baselineskip 

    \begin{subfigure}[b]{0.45\textwidth}
        \centering
        \includegraphics[width=0.65\textwidth]{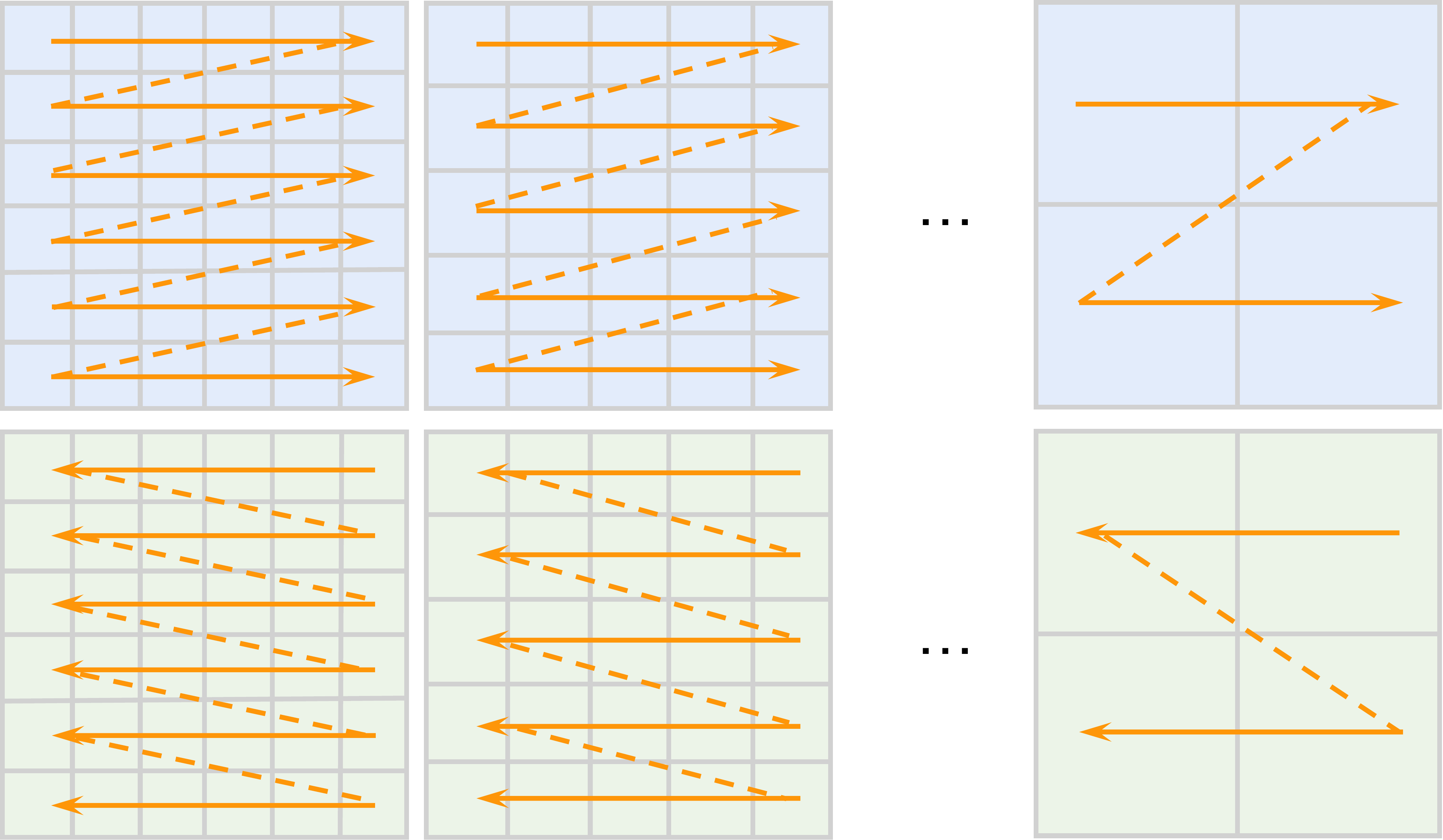}
        \caption{Hierarchical Scan~\cite{zhang2024motion}}
        \label{fig:hieratchical}
    \end{subfigure}
    \hfill
    \begin{subfigure}[b]{0.45\textwidth}
        \centering
        \includegraphics[width=0.85\textwidth]{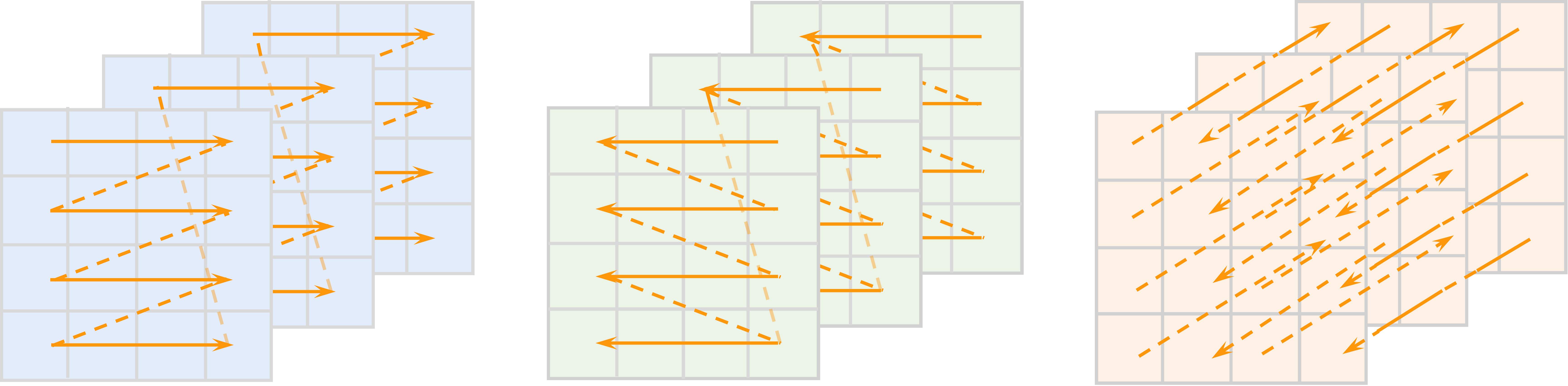}
        \caption{Spatiotemporal Selective Scan~\cite{yang2024vivim}}
        \label{fig:spatiotemporal}
    \end{subfigure}

    \vskip\baselineskip 

    \begin{subfigure}[b]{0.55\textwidth}
        \centering
        \includegraphics[width=0.75\textwidth]{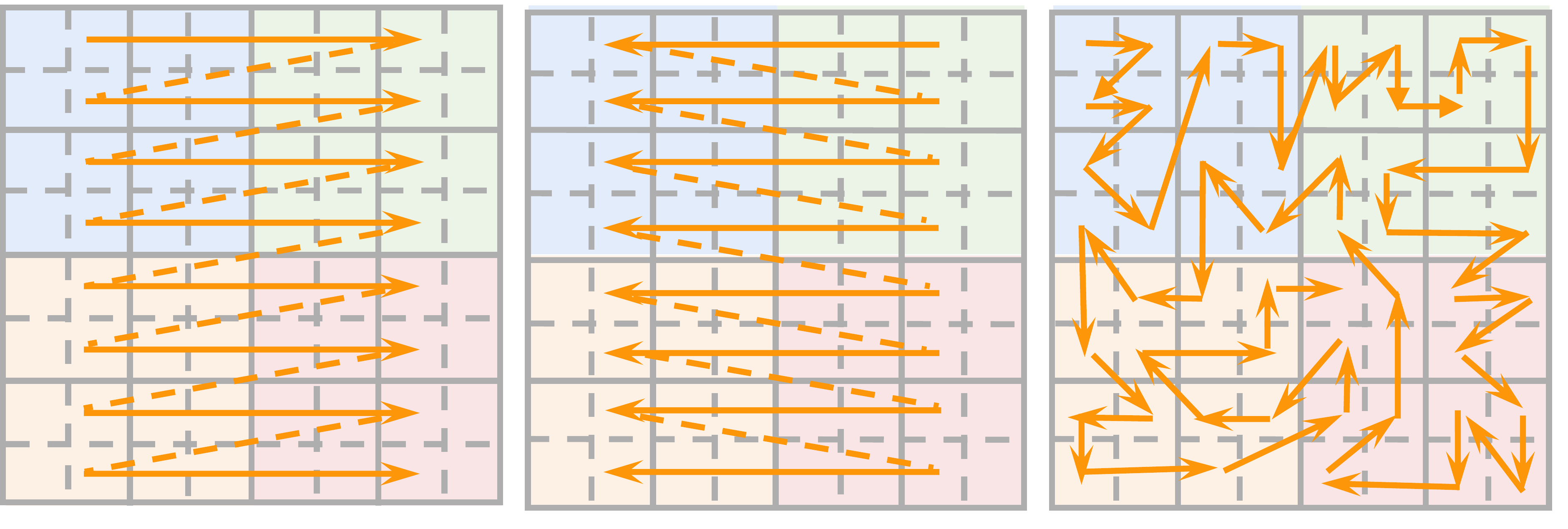}
        \caption{Multi-Path Scan~\cite{chen2024rsmamba}}
        \label{fig:multi-path}
    \end{subfigure}
    \caption{Comparison of various 2D scanning and selective scan orders in Vim, VMamba, PlainMamba, LocalMamba, Efficient VMamba, Zigzag, VMambaIR, VideoMamba, Motion Mamba, Vivim, and RSMamba. From~\cite{zhang2024survey}}
    \label{fig:scan-main}
\end{figure}

Following the introduction of Mamba, several researchers have explored modifications to this architecture, aiming
to enhance its performance and extend its applications \cite{behrouz2024mambamixer,sharma2024locating,wang2024mamba,chu2024incorporating,du2024understanding,pei2024efficientvmamba,liu2024mamba4rec,ali2024hidden,ezoe2024learning,fei2024scalable,agarwal2023spectral,wang2024mambabyte, shi2024vmambair,li2024videomamba, zhang2024motion}. This section reviews recent papers proposing various Mamba model architectural adaptations. Specifically, we delve into visual Mambas, which introduce diverse types of scanning mechanisms to understand and traverse direction-sensitive visual information. \autoref{fig:scan-main} presents a comprehensive overview of the landscape of scanning mechanisms within visual Mambas.

\textbf{Architecture Types}

Jamba \cite{lieber2024jamba} introduces a hybrid architecture that interleaves blocks of transformer and Mamba layers, integrating the strengths of both models. The use of a mixture-of-experts (MoE) approach within this framework allows Jamba to scale effectively while maintaining a manageable memory footprint. This model achieves SOTA performance on language benchmarks and long-context evaluations, handling up to 256K tokens in context length. Jamba's hybrid design showcases the potential of combining Mamba with other architectures to leverage their respective advantages.

Building on the concept of integrating MoE with mamba, MoE-Mamba \cite{pioro2024moe} significantly enhances the scalability and efficiency of SSMs. This architecture outperforms both the original Mamba and Transformer-MoE models, achieving the same performance as Mamba in 2.35× fewer training steps. MoE-Mamba's selective state space modelling and expert selection underscores the potential for efficient large-scale modelling in sequential tasks.

Blackmamba \cite{anthony2024blackmamba} takes the integration of MoE with Mamba a step further, combining the linear-complexity benefits of SSMs with the efficient inference capabilities of MoE. The model demonstrates competitive performance against both Mamba and transformer baselines, excelling in inference and training FLOPs. By open-sourcing its weights and inference code, Blackmamba encourages further exploration and application of this hybrid architecture in various domains.

Vision Mamba (Vim) \cite{zhu2024vision} addresses the challenge of visual representation learning by proposing a bidirectional Mamba block design. This architecture marks image sequences with position embeddings and compresses visual representation using bidirectional SSMs to effectively scan visual data (\autoref{fig:bidrectional}). Vim performs superior to established Vision Transformers (ViTs), such as DeiT~\cite{DeiT}, while significantly improving computation and memory efficiency. This positions Vim as a promising next-generation backbone for vision foundation models.

Visual Mamba (VMamba) introduces \cite{liu2024vmamba} a novel approach to visual representation learning by combining the strengths of CNNs and ViTs with the computational efficiency of SSMs. The architecture includes a Cross-Scan Module (CSM) to address direction-sensitive issues and traverse the spatial domain effectively, as illustrated in \autoref{fig:cross-scan}. Vmamba exhibits promising capabilities across various visual perception tasks, outperforming established benchmarks, particularly as image resolution increases.

ZigMa \cite{hu2024zigma} leverages Mamba's long sequence modelling capabilities for visual data generation, addressing scalability and complexity issues in diffusion models. The model introduces a Zigzag Mamba scan (\autoref{fig:zigzag}) and method that improves speed and memory utilization, outperforming Mamba-based and transformer-based baselines. ZigMa demonstrates its scalability on large-resolution visual datasets, showcasing its potential for efficient visual data generation.

Mamba-ND \cite{li2024mamba} extends the Mamba architecture to handle multi-dimensional data, such as images and videos. The model unravels input data across different dimensions in a row-major order, providing a systematic comparison with other multi-dimensional extensions. Mamba-ND shows competitive performance on benchmarks like ImageNet-1K~\cite{deng2009imagenet} classification and action recognition, demonstrating its versatility and efficiency in handling complex multi-dimensional data.

LocalMamba \cite{huang2024localmamba} addresses the limitations of Vim in capturing local dependencies within visual data by introducing a novel local scanning strategy (as shown in \autoref{fig:local-scan}). This approach divides images into distinct windows, effectively capturing local dependencies while maintaining a global perspective. Additionally, it employs a dynamic method to independently search for optimal scan choices for each layer, significantly improving performance. Extensive experiments demonstrate LocalMamba's ability to effectively capture image representations, outperforming Vim-Ti by 3.1\% on ImageNet~\cite{deng2009imagenet} with the same 1.5G FLOPs.

PlainMamba \cite{yang2024plainmamba} presents a simple non-hierarchical SSM designed for general visual recognition. This model enhances the selective scanning process of Mamba for visual data by employing a continuous 2D scanning process (\autoref{fig:cont2d}) that improves spatial continuity and a direction-aware updating mechanism. These innovations enable PlainMamba to learn features from two-dimensional images more effectively. The architecture is easy to use and scale, formed by stacking identical PlainMamba blocks with constant width throughout all layers. PlainMamba achieves performance gains over previous non-hierarchical models and is competitive with hierarchical alternatives, particularly excelling in tasks requiring high-resolution inputs while maintaining high performance with reduced computing requirements.

These papers represent diverse modifications and extensions to the Mamba architecture, each addressing specific challenges and applications. From hybrid models and visual representation learning to medical image segmentation and multi-dimensional data processing, the evolving landscape of Mamba-based models highlights the architecture's flexibility and potential for innovation. As researchers continue to explore and refine these models, the Mamba architecture is poised to play a crucial role in the future of deep learning.

\textbf{Clinical Importance}

The adoption of Mamba models in medical imaging marks a transformative advancement in clinical practice and research. These models bring several critical enhancements to traditional imaging techniques, aligning with the pivotal needs of modern medical diagnostics and therapeutic strategies.

\textbf{Enhanced Computational Efficiency:} One of the fundamental challenges in medical imaging is managing the extensive computational resources required for processing complex imaging data. With their linear complexity and hardware-aware algorithms, Mamba models facilitate rapid processing of large imaging datasets, such as full-body scans, which are common in routine clinical evaluations. This efficiency is crucial for real-time diagnostic applications, such as in emergency radiology, where quick decision-making can significantly impact patient outcomes.

\textbf{Extended Contextual Understanding:} Unlike conventional deep learning models that often require segmentation or region-specific annotations, Mamba models inherently understand and process extensive imaging data sequences. This ability is invaluable in medical imaging, where contiguous sections of a scan need to be evaluated in context to diagnose conditions such as cancers or vascular diseases accurately. For instance, the ability of Mamba to handle long data sequences can be leveraged to understand tumour progression over large areas or across multiple scans, enhancing both the diagnosis and monitoring phases.

\textbf{Reduced Radiation Exposure:} The efficient data processing capabilities of Mamba models allow for high-quality imaging results from lower-quality inputs \cite{zhang2024survey}. This feature can be particularly beneficial in reducing the radiation doses required in imaging techniques like CT scans. By needing fewer passes to produce diagnostic-quality images, Mamba models minimize the patient's exposure to potentially harmful radiation and reduce the operational costs associated with these procedures.

\textbf{Multi-modal Data Integration:} In clinical settings, the integration of various types of medical data (e.g., MRI, CT, PET) provides a more comprehensive view of a patient’s health status \cite{artesani2024empowering,azad2022medical}. Mamba models excel in processing and integrating diverse data types, offering a holistic view of the patient’s medical images. This capability ensures more accurate diagnoses and tailored treatment plans, improving patient care quality.

\textbf{Scalability to Advanced Applications:} Mamba models are designed to be scalable, a critical feature in medical research and clinical trials where voluminous data is processed. They can be adapted for advanced applications like genetic sequencing and biometric analyses, which are becoming increasingly relevant in personalized medicine and advanced diagnostics.

In summary, the Mamba model's ability to efficiently process extensive sequences, integrate multimodal data, and provide scalable solutions marks a significant advancement in the clinical application of medical imaging technologies. Its introduction into the medical field promises to enhance diagnostic accuracy, reduce operational inefficiencies, and pave the way for innovative therapeutic strategies, ultimately leading to improved patient care and health outcomes.

\section{State Space Models in Action}
\label{application}

\begin{figure}[!ht]
    \centering
    \includegraphics[width=0.84\textwidth]{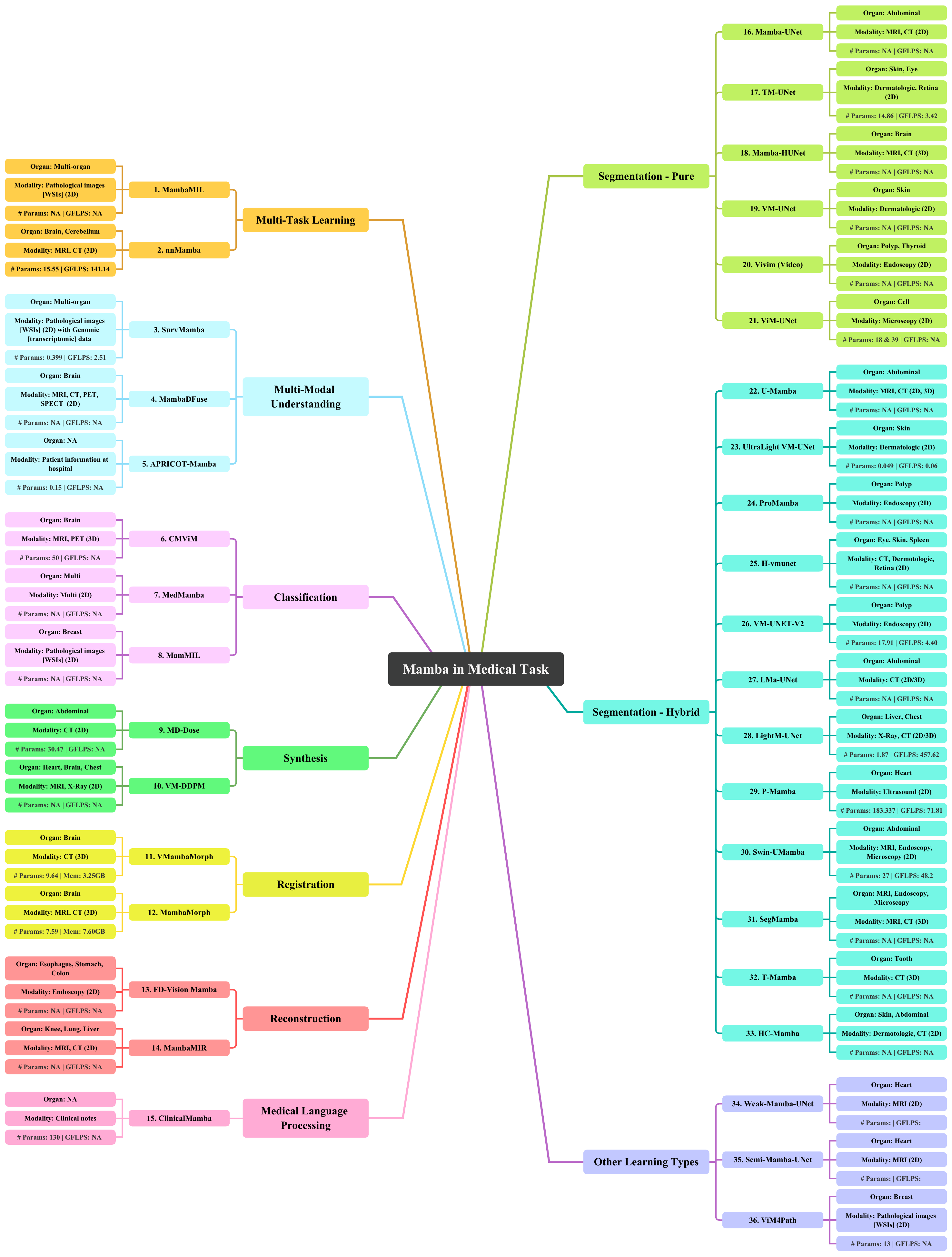}\caption{The proposed taxonomy for Mamba-based medical imaging research encompasses ten sub-fields: \Romannum{1}) Multi‑Task Learning, \Romannum{2}) Multi‑Modal Understanding, \Romannum{3}) Medical Image Classification, \Romannum{4}) Medical Image Synthesis, \Romannum{5}) Medical Image Registration, \Romannum{6}) Medical Image Reconstruction, \Romannum{7}) Medical Language Processing, and Medical Image Segmentation, which is subsequently divided into \Romannum{8}) Pure, \Romannum{9}) Hybrid and \Romannum{10}) Multi-disciplinary learning methods, referred to as other learning types. For conciseness, we use ascending prefix numbers in the paper’s title and denote the reference for each study accordingly: \protect\input{extras/taxonomy-refcs}}
    \label{fig:taxonomy}
\end{figure}

\subsection{Medical Image Segmentation}
\label{sec:segmentation}

Medical image segmentation is pivotal in various clinical applications, including disease diagnosis, treatment planning, and patient monitoring \cite{azad2023loss}. Accurate segmentation of anatomical structures and pathological regions from medical images is essential for extracting quantitative measurements, identifying abnormalities, and facilitating automated analysis. Selective SSMs offer a promising approach for medical image segmentation due to their ability to model temporal dependencies, handle long sequences of data, and capture complex spatial relationships within images while maintaining linear computational complexity compared to the quadratic burden of transformers and the small receptive fields of CNNs. Leveraging Mamba with selective state space mechanisms and efficient hardware-aware designs, Mamba-based models effectively model the complex structures and variations present in medical images. Integrating Mamba in medical image segmentation tasks enhances segmentation accuracy and reduces the need for large-scale datasets and massive computational resources, making it a valuable tool in resource-constrained environments.

In this context, we categorize Mamba-based segmentation networks into three groups based on their initial structure and training methodology. The taxonomy includes: \textit{"Pure SSM and Mamba-based Models (\textbf{Denoted as Pure)}"} which mainly utilize visual SSM modules for segmentation tasks, exploiting their ability to model long-range dependencies and capture complex spatial relationships within medical images. Additionally, \textit{"Hybrid Architectures Involving Mamba (\textbf{Denoted as Hybrid)})"} approaches incorporate convolutional layers alongside SSM modules to enhance feature capture and improve local representation, combining the strengths of both techniques for more effective segmentation results. Finally, \textit{"Other Learning Types"} approaches use alternative learning techniques such as self-supervised, semi-supervised, and weakly supervised learning to boost segmentation performance.

\subsubsection{Pure SSM and Mamba-based Models}

Wang et al. \cite{wang2024mamba} introduced Mamba-UNet, a novel architecture that combines the UNet in medical image segmentation with Mamba's capabilities. Mamba-UNet adopts a VMamba-based encoder-decoder structure, incorporating skip connections to retain spatial information across different network scales. This design facilitates comprehensive feature learning, capturing intricate details and broader semantic contexts within medical images. A novel integration mechanism within the VMamba blocks ensures seamless connectivity and information flow between the encoder and decoder paths, enhancing segmentation performance.
TM-UNet~\cite{tang2024rotate} is introduced to address the challenge that visual SSMs feature extraction methods often retain redundant structures, leaving room for parameter reduction. Featuring Residual VSS Blocks and a Triplet SSM Module, TM-UNet improves information transfer, mitigates network degradation, and effectively combines spatial and channel features through the Triplet SSM, avoiding the extra parameters of conventional attention mechanisms. Mamba-HUNet~\cite{sanjid2024integrating} is another architecture that captures localized fine-grained features and long-range dependencies while ensuring linear feature size scaling, avoiding the quadratic complexity often seen in transformers in purely SSM-based U-shaped networks. Another UNet structure in cooperate with VMamba is VM-UNet ~\cite{ruan2024vmunet} which presents a pure SSM-based model developed to demonstrate its potential in medical image segmentation tasks. In this model, the Visual State Space (VSS) block is designed to effectively capture extensive contextual information by utilizing an asymmetrical encoder-decoder structure, making it a foundational component in the system. Moreover, ViM-UNet~\cite{archit2024vimunet} is proposed as a model for microscopy instance segmentation based on visual Mamba.

One of the pioneering efforts to integrate SSMs into medical video object segmentation, resulting in enhanced speed and performance, is Vivim ~\cite{yang2024vivim}. This is a framework that integrates Mamba into a multi-level transformer architecture for medical video object segmentation. The authors have designed a novel Temporal Mamba Block and incorporated a structured state-space sequence model with a spatiotemporal selective scan, ST-Mamba, to enhance video visual representation learning. Additionally, they have employed a boundary-aware constraint and collected a thyroid segmentation dataset VTUS for benchmarking. Their experiments validate the effectiveness of Vivim. The Vivim system consists of a hierarchical encoder with stacked Temporal Mamba Blocks to extract feature sequences and a lightweight CNN-based decoder head to make predictions for segmentation masks. 

\subsubsection{Hybrid Architectures Involving Mamba}

\begin{figure}[hbt]
    \centering
    \includegraphics[width= 1\textwidth]{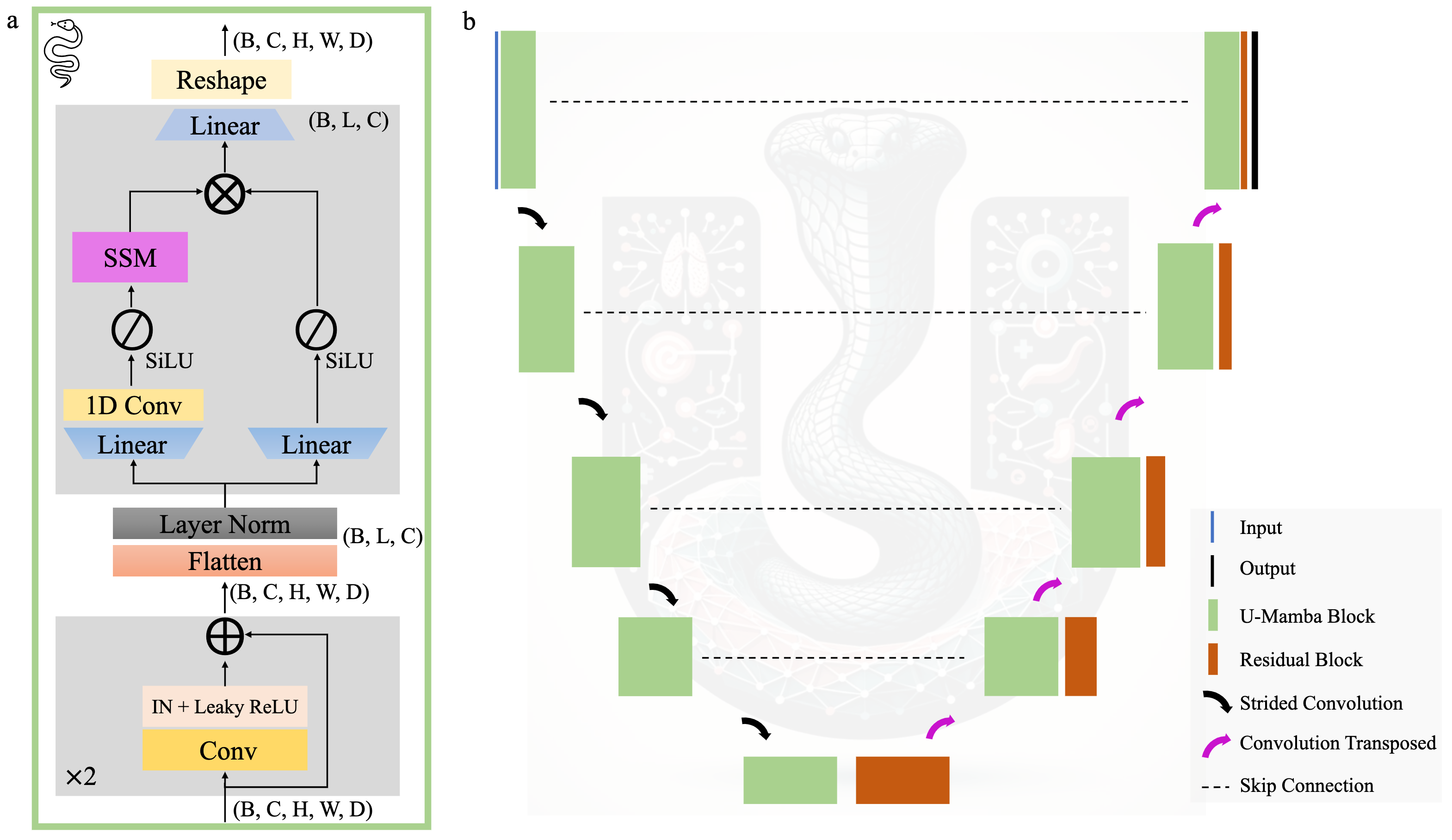}  \caption{Overview of the U-Mamba architecture: a) Each U-Mamba block has two Residual blocks followed by an SSM-based Mamba block for long-range dependency modelling. b) The encoder-decoder framework includes U-Mamba blocks in the encoder, Residual blocks in the decoder, and skip connections. From~\cite{ma2024umamba}.}
    \label{fig:umamba}
\end{figure}

U-Mamba~\cite{ma2024umamba} combines CNNs and SSMs to form a hybrid framework. This blend capitalizes on CNNs' proficiency in extracting local features and SSMs' ability to grasp extensive relationships within images. Structured with an encoder-decoder setup, as illustrated in \autoref{fig:umamba}, the architecture boosts its effectiveness in handling long-range data and adapts well to diverse segmentation assignments. Through the fusion of these elements, U-Mamba surpasses conventional CNN and Transformer-based models in medical image segmentation tasks, showcasing enhanced performance.
In an exploration of the potency of pretraining, which has demonstrated significant efficacy in data-efficient medical image analysis, Swin-UMamba ~\cite{liu2024swinumamba} is designed specifically for medical image segmentation tasks, harnessing the benefits of ImageNet-based pretraining.

To improve the Mamba model, Vision Mamba UNetV2 (VM-UNetV2) ~\cite{zhang2024vmunetv2} reintegrates both low-level and high-level features. It infuses semantic information into the low-level features and refines the high-level features using more detailed information. SegMamba~\cite{xing2024segmamba} also is an architecture designed for 3D medical image segmentation, combining a U-shaped structure with Mamba to capture global features across entire volumes at multiple scales. To optimize Mamba for high-dimensional medical images, the architecture includes a tri-orientated Mamba (ToM) module that enhances the sequential modelling of 3D features from three different directions. A gated spatial convolution (GSC) module is incorporated to improve spatial feature representation before each ToM module.
In another work, High-order Vision Mamba UNet (H-VMUnet) ~\cite{wu2024hvmunet} proposes a High-order 2D-selective-scan (HSS2D) to guarantee an outstanding global sensory field within SS2D, while also reducing redundant information.
Wang et al.~\cite{wang2024large} introduce the LMa-UNet, a pioneering solution for segmenting 2D and 3D medical images. Its innovation lies in the use of large windows, transforming local spatial modelling compared to traditional methods. Remarkably efficient in global modelling, it outperforms self-attention mechanisms. Additionally, a novel hierarchical and bidirectional Mamba block enhances both global modelling capabilities and proficiency in capturing intricate spatial relationships.
Introducing the polyp segmentation model, Prompt-Mamba ~\cite{xie2024promamba} integrates Vision-Mamba as its backbone while implementing prompt assistance, showcasing superior generalization performance on downstream datasets. Also, Ye et al.~\cite{ye2024pmamba} introduce P-Mamba for efficient pediatric echocardiographic left ventricular segmentation. P-Mamba comprises three key parts: the Vim encoder branch, the DWT-based PMD encoder branch, and decoders. Each branch generates four feature maps, starting from higher resolutions and gradually decreasing while also augmenting the channels. The Vim branch is responsible for ensuring the model's efficiency and capturing overarching dependencies, while the DWT-based PMD branch focuses on eliminating noise interference and preserving the integrity of target edges to capture localized features. During the decoding process, the outputs from both branches are merged and processed through Segmentation Head (SegHead) and Fully Convolutional Network Head (FCNHead) decoders, which generate segmentation masks by increasing resolution. Each decoder is guided by a specific loss function to enhance the overall performance of the network.
Moreover, T-Mamba ~\cite{hao2024tmamba} presents a novel approach by integrating shared positional encoding and frequency-based features into Vim, effectively addressing concerns regarding spatial position preservation and feature enhancement within the frequency domain. Additionally, it incorporates a gate selection unit that dynamically combines two spatial domain features with one frequency domain feature. T-Mamba marks the pioneering integration of frequency-based features into Vim.

Various approaches have been proposed to reduce computational complexity, such as Light M-UNet ~\cite{liao2024lightmunet} and UltraLight VM-UNet~\cite{wu2024ultralight}. UltraLight VM-UNet ~\cite{wu2024ultralight} utilizes a Parallel Vision Mamba (PVM) layer, an approach within Vision Mamba, which processes features in parallel. This method attains outstanding performance with minimal computational complexity while maintaining a consistent number of processing channels.  UltraLight VM-UNet has a total of 6-layer structure consisting of a U-shaped structure. From layers 4 to 6, each layer incorporates a PVM Layer. The decoder section also replicates the encoder's configuration. The skip-connection pathway employs both the Channel Attention Bridge (CAB) module and the Spatial Attention Bridge (SAB) module to fuse multi-level and multi-scale information. In another approach, the HC-Mamba~\cite{xu2024hcmamba}\ model introduces dilated convolution to capture a wider range of contextual details without increasing computational expenses. Furthermore, HC-Mamba employs depthwise separable convolutions, decreasing parameters and computational demands. This fusion of methods enables HC-Mamba to efficiently handle extensive medical image datasets with reduced computational overhead while upholding excellent performance.



\subsubsection{Other Learning Types}


The current paradigm of training deep learning models necessitates vast amounts of labelled training data. This process is not only time-intensive but also costly, particularly when it comes to gathering medical images, prompting the path for methods based on limited supervision \cite{huang2023self,peng2021medical}. These challenges have inspired Mamba-based research efforts on
learning medical image segmentation with limited supervision. In this regard, Vim4Path \cite{nasirisarvi2024vim4path} adopts a modified version of the Vim \cite{zhu2024vision} architecture with positional encoding \cite{dosovitskiy2020image} within the seminal DINO \cite{caron2021emerging} network to sequentially process on patch level from
the Gigapixel Whole Slide Image
(WSI). Specifically, Vim4Path exploits a self-distillation setup, where two Vim networks are considered teacher-student, guiding the student’s
training without labelled data. To compensate for the mismatch of the dimensions of the input image, bicubic interpolation is employed to modify the patch embeddings based on the target grid size, which is determined from the input dimensions and patch size. This method simulates the way cancer cells are realistically represented, transitioning from close-up to a broader view on a slide, similar to how pathologists navigate slides under a microscope to diagnose cancer cells.
Two related studies \cite{wang2024weakmambaunet, ma2024semimambaunet} adopt the Mamba architecture to enhance model efficacy with sparse annotations with distinct supervisory methods. In particular, The Weak-Mamba-UNet \cite{wang2024weakmambaunet} integrates CNN, ViT, and Mamba architectures for scribble-based medical image segmentation. This method employs a collaborative cross-supervisory mechanism that uses pseudo-labels to enhance learning and refinement across the three architectures. Similarly, Semi-Mamba-UNet \cite{ma2024semimambaunet} merges a conventional CNN-based UNet with a Mamba-based U-shaped encoder-decoder within a semi-supervised learning framework.
It introduces a pixel-level contrastive learning strategy that maximizes feature extraction from labelled and unlabeled data.
Both methods demonstrate superior performance on the ACDC MRI cardiac segmentation dataset \cite{bernard2018deep} by effectively leveraging sparse annotations through an integrated approach, demonstrating the range of limited-supervised learning contexts that benefit from the unique features of the visual Mamba.

\subsection{Medical Image Classification}

\begin{figure}[hbt]
    \centering
    \includegraphics[width= 0.75\textwidth]{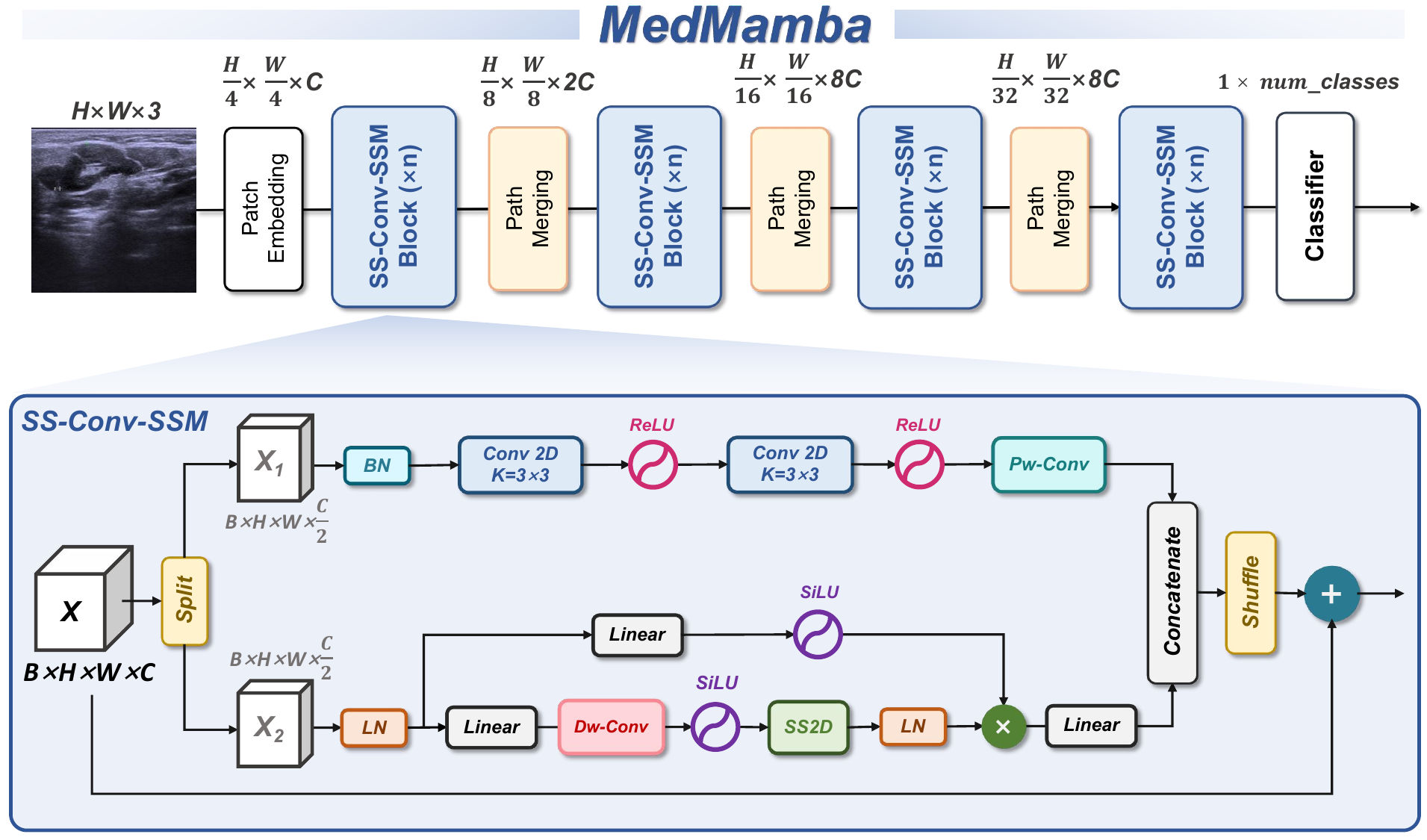}    
    \caption{MedMamba architecture and illustration of SS-Conv-SSM block. From~\cite{yang2024clinicalmamba}}
    \label{fig:medmamba}
\end{figure}

Medical image classification is a fundamental and essential task in the field of CV. MedMamba architecture~\cite{yue2024medmamba}, inspired by the success of VMamba in natural image classification, incorporates the SS-Conv-SSM module to enhance feature extraction capabilities. This module uniquely combines convolutional layers, which capture local features, with SSMs for long-range modelling, as illustrated in \Cref{fig:medmamba}. This integration enables MedMamba to classify medical images across various datasets effectively. The SS-Conv-SSM Block, a central component of MedMamba, adopts a dual-branch approach using a 2D-selective-scan (SS2D) to ensure robust integration of both fine-grained and coarse-grained features, thus maximizing classification efficacy.

Yang et al.~\cite{yang2024cmvim} introduce the Contrastive Masked Vim Autoencoder (CMViM), a novel representation learning method designed for 3D multi-modal data. This approach incorporates the Vim into the masked autoencoder framework, effectively reconstructing and modelling long-range dependencies in 3D medical data for Alzheimer's disease (AD) classification. The method consists of two key steps: firstly, leveraging Vim within the masked autoencoder to enhance the modelling of 3D masked multi-modal data, and secondly, applying contrastive learning mechanisms to align multi-modal representations from both intra-modal and inter-modal perspectives. This dual strategy enhances the discriminative capabilities of image representations and ensures effective alignment of multi-modal data, integrating intra-modal learning from the same modality and inter-modal learning across different modalities to optimize representation learning for AD classification.

Furthermore, MamMIL~\cite{fang2024mammil} integrates Mamba into multiple instance learning (MIL), achieving efficient WSI classification. The model features a bidirectional SSM and a 2D content-aware block using pyramid-structured convolutions to learn bidirectional instance dependencies and 2D spatial relations. MamMIL demonstrates advanced classification performance and uses less GPU memory than transformer-based methods.

\subsection{Medical Image Synthesis} 

\begin{figure}[!tbh]
    \centering
    \includegraphics[scale = 0.55]{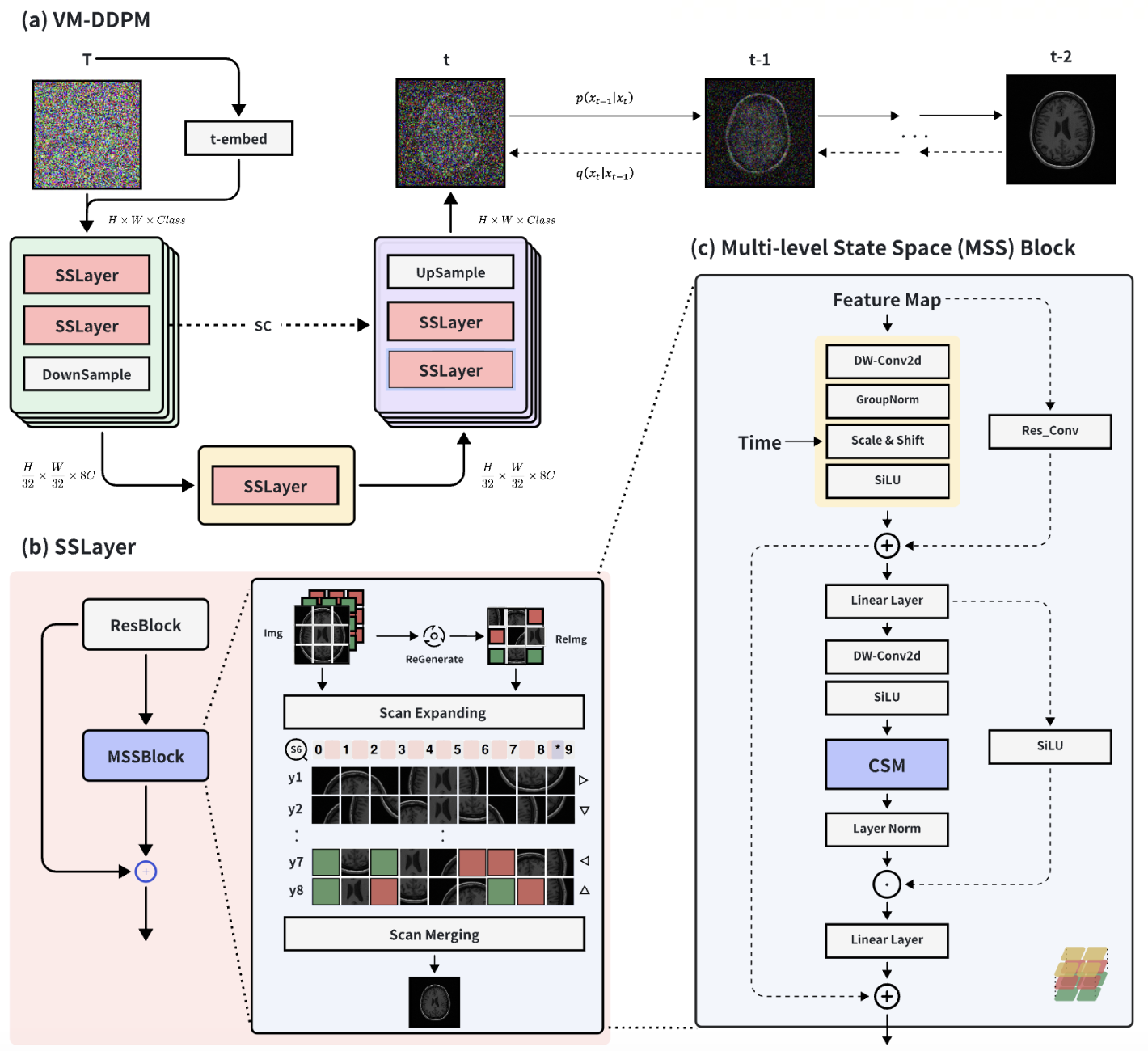} 
    \caption{(a) The general structure of VM-DDPM. (b) The primary building layer of VM-DDPM is the SSLayer. (c) The central element of SSLayer is the 
    MSSBlock incorporates time embedding and CSM operations. From~\cite{ju2024vmddpm}}
    \label{fig:vmddpm}
\end{figure}

In advancing smart healthcare, researchers are improving the scale and diversity of medical datasets through medical image synthesis. Existing methods face limitations due to CNN's local perception and the quadratic complexity of transformers, complicating the balance of structural texture consistency. To address these challenges, the Vision Mamba DDPM (VM-DDPM)~\cite{ju2024vmddpm} has been proposed. VM-DDPM is based on a SSM that integrates CNN's local perception with SSM's global modeling capabilities(as illustrated in \Cref{fig:vmddpm}), maintaining linear computational complexity. The design includes a multi-level feature extraction module, the Multi-level State Space Block (MSSBlock), and an encoder-decoder structure's basic unit, the State Space Layer (SSLayer), specifically for medical pathological images. Additionally, a Sequence Regeneration strategy enhances the Cross-Scan Module (CSM), enabling the S6 module to fully perceive 2D image spatial features and stimulate model generalization. This innovative approach marks the first application of the SSM-CNN hybrid architecture in medical image synthesis, demonstrating competitive performance on various datasets, along with qualitative assessments by radiologists.

In a related effort, MD-Dose~\cite{fu2024mddose} introduces an innovative diffusion model based on the Mamba architecture for predicting radiation therapy dose distribution in thoracic cancer patients. MD-Dose adds Gaussian noise to dose distribution maps in the forward process to create pure noise images and uses a Mamba-based noise predictor to reconstruct the dose distribution maps in the backward process. A Mamba encoder extracts structural information integrated into the noise predictor to localize dose regions within the planning target volume (PTV) and organs at risk (OARs). Experiments on a dataset of 300 thoracic tumour patients demonstrate MD-Dose's superior performance across various metrics and in terms of time efficiency.

\subsection{Medial Image Registration}


Capturing voxel-wise spatial correspondence across distinct modalities is essential for medical image registration~\cite{Epilepsy}, yet current registration methods often lack accuracy and clinical practicality. MambaMorph~\cite{guo2024mambamorph}, an innovative multi-modality deformable registration framework, leverages a Mamba-based module for efficient long-range spatial modelling and a fine-grained UNet-inspired feature extractor for high-dimensional learning. Featuring well-aligned MR-CT volume pairs, MambaMorph excels in both registration accuracy and speed, outperforming existing methods and highlighting its significant potential for practical medical applications.

VMambaMorph~\cite{wang2024vmambamorph}, a groundbreaking hybrid VMamba-CNN network, is designed for 3D image registration. VMambaMorph computes the deformation field based on target and source volumes using a U-shaped network architecture. The VMamba-based block, redesigned with a 2D cross-scan module for 3D volumetric feature processing and a fine-grained feature extraction module, is proposed for high-dimensional feature learning. VMambaMorph incorporates a fine-grained feature extractor to achieve high-level spatial features more effectively.

\subsection{Medial Image Reconstruction}

In endoscopic imaging, maintaining high-quality images is crucial for assisting healthcare professionals in decision-making, as recorded images are prone to exposure abnormalities~\cite{sang2020inferring}. To address this issue, motivated by the success of Mamba in visual resotoration~\cite{huang2024mambamir}, FDVision Mamba (FDVM-Net)~\cite{zheng2024fdvision}, a frequency-domain based network, has been designed to achieve high-quality image exposure correction by reconstructing the frequency domain of endoscopic images. Inspired by SSMs, FDVM-Net incorporates a C-SSM block that combines convolutional layers' local feature extraction ability with the SSM's capability to capture long-range dependencies. The network features a dual-path design, where each path processes the phase and amplitude information of the image separately. Additionally, a novel frequency-domain cross-attention mechanism enhances model performance by determining which information values to preserve or overlook, thereby improving image quality and supporting better clinical decision-making.

    



Similarly, MambaMIR~\cite{huang2024mambamir}, a Mamba-based model for medical image reconstruction, and its Generative Adversarial Network-based variant, MambaMIR-GAN, are introduced. MambaMIR inherits several advantages from the original Mamba model, such as linear complexity, global receptive fields, and dynamic weights. These features are particularly well-suited for medical image reconstruction tasks, which require processing long sequences (large spatial resolutions) to preserve detailed information and ensure global sensitivity through long-range dependency. A novel arbitrary-mask mechanism is introduced to adapt Mamba to this task, providing randomness for subsequent Monte Carlo-based uncertainty estimation by randomly masking out redundant image scan sequences. This mechanism enhances the model's ability to handle variability in medical imaging data, making MambaMIR and MambaMIR-GAN powerful tools for both image reconstruction and uncertainty estimation.

\subsection{Medical Language Processing}
NLP systems have enormous potential in healthcare due to their ability to comprehend complex clinical notes. The paper, ~\cite{yang2024clinicalmamba}, introduces a language model designed to analyze longitudinal clinical notes with unmatched precision, understand clinical context, and predict future patient outcomes. ClinicalMamba harnesses the power of deep learning architecture trained on a vast dataset of electronic health records, generating coherent and relevant clinical text that has significant implications in healthcare. The proposed model captures temporal relationships and context within patient records.
\begin{figure}[hbt]
    \centering
    \includegraphics[scale = 0.39]{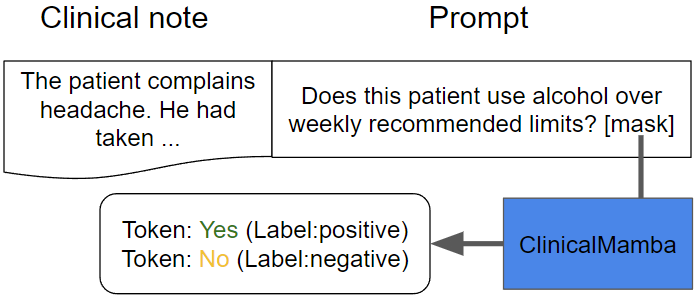}    
    \caption{Prompt-based fine-tuning ~\cite{yang2024clinicalmamba}}
    \label{fig:cmlp}
\end{figure}

The provided text outlines a three-step methodology for training and fine-tuning a language model tailored for clinical NLP tasks with limited labelled data. The methodology involves pre-training on MIMIC-III clinical notes \cite{mimic-iii}, prompt-based fine-tuning (illustrated in \Cref{fig:cmlp}, and fine-tuning for two specific clinical tasks: cohort selection for clinical trials and ICD coding. The researchers also leveraged previous discharge summaries and incorporated ICD code descriptions from prior visits to enhance the quality and effectiveness of their approach.

\subsection{Multi-Modal Understanding}

Multi-modal learning, which combines pathological images with genomic data, has significantly enhanced the accuracy of survival prediction in cancer patients. This approach leverages the advantages of both data types; pathological images detail the tumour microenvironment, capturing the diversity of cancer cells and immune interactions~\cite{GUI2023e515}, while genomic profiles provide critical insights into cancer cell states and immune system factors that affect dynamic prognostic outcomes~\cite{tong2023prioritizing}. By integrating these complementary sources of information, multi-modal learning holds considerable promise for improving the precision of cancer survival predictions.

SurvMamba \cite{chen2024survmamba} leverages state space modelling and multi-grained multi-modal interactions to enhance its predictive performance, addressing the limitations of traditional models. The paper introduces the Mamba model for multi-modal survival prediction. It efficiently processes high-dimensional WSIs and transcriptomic data, outperforming various SOTA methods with lower computational cost. The proposed Hierarchical Interaction Mamba (HIM) module encodes comprehensive intra-modal representations at fine-grained and coarse-grained levels from WSI and transcriptomic data. An Interaction Fusion Mamba (IFM) module also integrates histological and genomic features across various levels, capturing multi-modal features from diverse perspectives. The system creates a "bag" of each WSI and transcriptomics data, extracting features of small instances and using a bidirectional Mamba in a dual-level Multiple Instance Learning (MIL) framework to combine the small instances into larger ones. The IFM module enables interactions between histological and genomic features to derive representations at both small and large scales. Finally, these multi-scale representations are fused to predict a hazard function and obtain precise survival risk scores with confidence.
 
The MambaDFuse model, presented in \cite{li2024mambadfuse}, is designed for multi-modality image fusion (MMIF). It operates in three phases: feature extraction, fusion, and reconstruction. The model preserves essential information from different modalities, such as MRI and CT scans, to produce high-quality fused images for medical imaging applications.  Each modality's image is processed separately to extract relevant features in the feature extraction phase. In the fusion phase, the extracted features are combined using a fusion mechanism to generate a fused representation that preserves essential information from both modalities. In the final step, the fused image reconstruction module applies the inverse transformation of the feature extraction process to produce a high-quality fused image. The proposed model effectively preserves details and enhances the quality of fused images, making it valuable for various medical imaging applications, including diagnosis and treatment planning.
\begin{figure}[hbt]
    \centering
    \includegraphics[width= 1\textwidth]{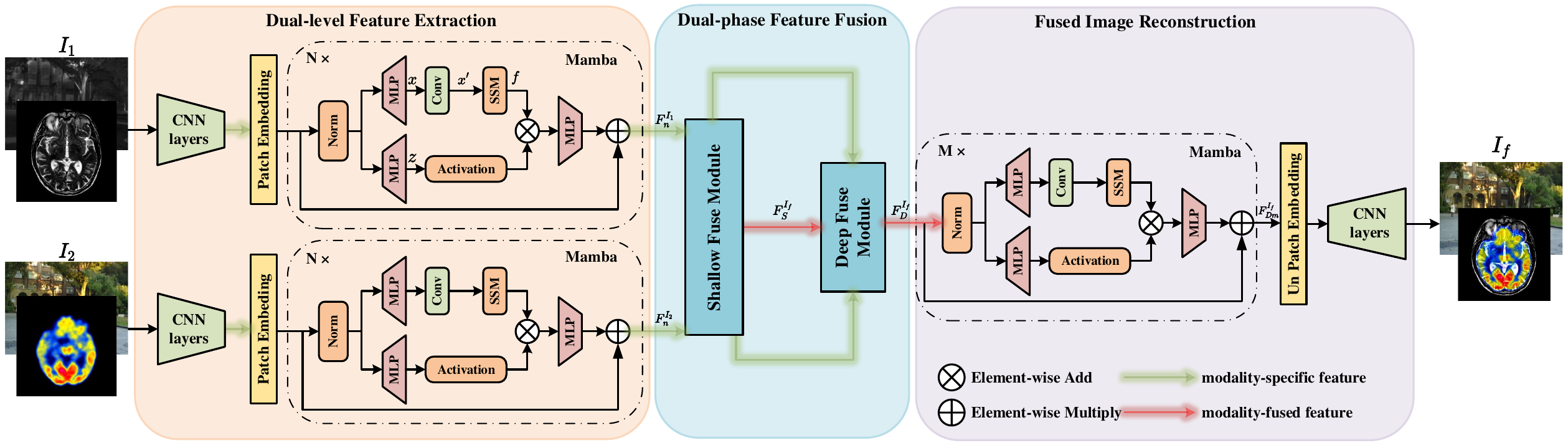}  \caption{The architecture of MambaDFuse \cite{li2024mambadfuse}. The system comprises three stages: dual-level feature extraction, dual-phase feature fusion, and fused image reconstruction.}
    \label{fig:mambadfusea}
\end{figure}
The research focuses on enhancing image fusion techniques with MambaDFuse (\Cref{fig:mambadfusea}), a more effective method than traditional CNNs and transformers. The proposed dual-level feature extractor and dual-phase feature fusion module achieve SOTA image fusion performance for both Infrared-visible image fusion (IVF) and Multi-modality image fusion (MIF). The method's fusion images effectively facilitate downstream object detection, indicating potential for future advancements in the field of image fusion.

A prediction model for acuity in the Intensive Care Unit (ICU) named Stability, Transitions, and Life-Sustaining Therapies Prediction Model is presented in \cite{contreras2024apricotmamba}. The model, developed and validated in the paper, aims to forecast the clinical condition of patients in the ICU, focusing on stability, transitions, and necessary life-sustaining therapies. Leveraging the Mamba framework, the model utilizes a combination of deep learning techniques and clinical data to make accurate predictions regarding patient outcomes. Through rigorous development and validation processes, Acuity Prediction in Intensive Care Unit-Mamba (APRICOT-Mamba) demonstrates promising performance in predicting ICU patient acuity, offering potential benefits for clinical decision-making and patient care management in critical care settings. APRICOT-M is an algorithm that accurately predicts the probability of a patient being stable or unstable, along with the probability of mortality or discharge. It also predicts transitions between acuity states and the need for life-sustaining therapies, such as mechanical ventilation (MV), vasopressors (VP), and continuous renal replacement therapy (CRRT). 

\subsection{Multi Task Learning}\label{sec:multi-task}
Multi-task learning offers significant advantages and is essential for efficiently leveraging shared representations and improving the generalization capabilities of models across various tasks. By learning multiple related tasks simultaneously, models can capture commonalities and differences between tasks, leading to better performance and reduced risk of overfitting. This approach is particularly beneficial in complex domains such as 3D medical imaging, where tasks like segmentation, classification, and landmark detection are interrelated.

Integrating Mamba blocks in CNNs exemplifies an innovative approach in 3D medical imaging, as demonstrated by nnMamba \cite{gong2024nnmamba}. This led to the creation of nnMamba, a versatile architecture that sets a new standard for segmentation, classification, and landmark detection. The authors implemented a UNet architecture, strategically incorporating the Mamba layer early to empower features with global context from the start.

The paper introduces Mamba as a parallel module to further enhance the CNNs' capability for modelling long-range dependencies. Additionally, the paper proposes the mamba-In-Convolution (MIC) block, which combines SSM with Convolutional layers, including a 1 × 1 convolution, batch normalization, and ReLU activation. The paper introduces the Channel-Spatial Siamese (CSS) input module, which fully harnesses the representational capabilities of SSMs, inspired by Siamese Networks \cite{siamese, unbiased}.
It also emphasizes the effectiveness of the Mamba-In-Convolution with Channel-Spatial Siamese input (MICCSS) module for long-range relationship modelling. The authors improved performance in dense prediction tasks by using an encoder with the MICCSS module and skip scaling. Additionally, they enhanced the stem layer with the MICCSS module for classification tasks, leading to significant improvements in accuracy and performance.

In \cite{yang2024mambamil}, a new MambaMIL model is proposed to address challenges in Multiple Instance Learning (MIL). It includes the Reordering Mamba (SR-Mamba) component, which captures long-range dependencies among positive instances. This model has shown promising results in computational pathology tasks, such as cancer diagnosis and tissue classification, highlighting the effectiveness of sequence reordering for image analysis. With MambaMIL, each instance can interact with any previously scanned instances through a compressed hidden state, making it easier to model long sequences while mitigating computational complexity.
The authors partition the tissue regions of a WSI into patches and map them into instance features using a feature extractor. The features are then passed through linear projection and a series of stacked SR-Mamba modules to model long sequences. Finally, an aggregation module obtains bag-level representations for downstream tasks, capturing comprehensive contextual information with minimal computational complexity.

\section{Discussion}
\label{sec:analysis}
{\renewcommand{\arraystretch}{1.25}%

\begin{table}[!ht]
\centering
\caption{Overview of the assessed Mamba-based models in medical applications, emphasizing their core ideas and notable highlights.}
\resizebox{\textwidth}{!}{
    \begin{tabular}{ >{\raggedright}p{4cm} >{\raggedright}p{4cm} > {}p{8cm} > {}p{8cm} }
    \bottomrule
    \rowcolor{black!5} \textbf{Application} & \textbf{Paper Names} & \textbf{Core Idea} & \textbf{Highlights} \\ \bottomrule
    \rowcolor{orange!10} \textbf{Multi-Task Learning} & MambaMIL~\cite{yang2024mambamil} \\ nnMamba~\cite{gong2024nnmamba} & Utilizing shared representations enhances generalization and reduces overfitting, improving segmentation, classification, and landmark detection in 3D medical imaging. Mamba-CNN blocks effectively capture long-range dependencies and sequence relationships for vital medical tasks such as cancer diagnosis. & $\bullet$ Enhancing dense prediction versatility across multiple tasks~\cite{gong2024nnmamba}\newline $\bullet$ Using Channel-Spatial Siamese modules for improved accuracy~\cite{gong2024nnmamba}\newline $\bullet$ Maintains minimal computational complexity for efficiency~\cite{gong2024nnmamba, yang2024mambamil}\newline $\bullet$ Employs shared representation to optimize learning~\cite{yang2024mambamil}\\ \hline
    
    \rowcolor{cyan!10} \textbf{Multi-Modal Understanding} & SurvMamba~\cite{chen2024survmamba} \\ MambaDFuse~\cite{li2024mambadfuse} \\ APRICOT-Mamba~\cite{contreras2024apricotmamba} & Combining pathological images and genomic data improves survival prediction by leveraging their strengths. This method uses state space modeling and multi-modal interactions for efficient processing and high-quality image fusion. Deep learning with clinical data predicts ICU patient acuity, stability, transitions, and life-sustaining therapies &  $\bullet$ Producing high-quality fused images from multiple sources~\cite{li2024mambadfuse}\newline $\bullet$ Enhances understanding through multi-modal interactions and achieving effective multi-modality image fusion techniques~\cite{chen2024survmamba, li2024mambadfuse} \\ \hline
    
    \rowcolor{violet!10} \textbf{Medial Image Classification} & CMViM~\cite{yang2024cmvim} \\ MedMamba~\cite{yue2024medmamba} \\ MamMIL~\cite{fang2024mammil} & Combining convolutional layers with SSMs improves classification accuracy. Strategies like masked autoencoders and contrastive learning enhance 3D medical data classification. Multiple instance learning with bidirectional SSMs and pyramid-structured convolutions efficiently classifies whole slide images while reducing memory usage. & $\bullet$ Utilizing contrastive learning for improved feature differentiation~\cite{yang2024cmvim}; \newline $\bullet$ Employs multiple instance learning for robust classification~\cite{fang2024mammil}\newline $\bullet$ Ensures efficient GPU usage for faster processing~\cite{fang2024mammil} \\ \hline
    
    \rowcolor{green!10} \textbf{Medial Image Synthesis} & MD-Dose~\cite{fu2024mddose} \\ VM-DDPM~\cite{ju2024vmddpm} & Combining CNN's local perception with SSM's global modeling, this approach maintains linear computational complexity. The diffusion process refines image details through noise addition and removal. Mamba and diffusion models together enhance radiation therapy dose prediction and medical image synthesis by managing noise and extracting structural information effectively & $\bullet$ Leveraging sequence regeneration, enhance spatial feature perception and model generalization~\cite{ju2024vmddpm}\newline $\bullet$ Using a Mamba-based diffusion model, efficiently reconstruct dose distribution maps and image synthesis with high accuracy~\cite{ju2024vmddpm, fu2024mddose}\newline $\bullet$ Maintaining time efficiency by ensuring fast processing times while delivering high-quality data synthesis~\cite{fu2024mddose, ju2024vmddpm} \\ \hline
    
    \rowcolor{yellow!40} \textbf{Medial Image Registration} & VMambaMorph~\cite{wang2024vmambamorph} \\ MambaMorph~\cite{guo2024mambamorph} & Enhancing medical image registration involves capturing voxel-wise spatial correspondence through efficient long-range spatial modeling and high-dimensional feature learning. Using Mamba-based modules and a UNet-inspired feature extractor, these frameworks achieve superior registration accuracy and speed, making them practical for medical applications. & $\bullet$ Excelling in aligning MR-CT volume pairs with high accuracy~\cite{guo2024mambamorph}\newline $\bullet$ Providing high registration speed through high-dimensional feature learning~\cite{guo2024mambamorph}\newline $\bullet$ Efficient long-range spatial modeling enhances registration performance and  multi-modality deformable registration capabilities~\cite{wang2024vmambamorph}\newline $\bullet$ Capturing voxel-wise spatial correspondence for better accuracy~\cite{guo2024mambamorph}.\\ \hline
    
    \rowcolor{red!15} \textbf{Medial Image Reconstruction} & FD-Vision Mamba~\cite{zheng2024fdvision} \\ MambaMIR~\cite{huang2024mambamir} &  Utilizing a dual-path design with frequency-domain cross-attention to improve image quality. Linear complexity, global receptive fields, and dynamic weights preserve detailed information, while an arbitrary-mask mechanism handles variability and provides uncertainty estimation. & $\bullet$ Enhancing image quality via frequency-spatial dual-path reconstruction~\cite{zheng2024fdvision}\newline $\bullet$ Cross-attention with linear complexity preserves detailed information~\cite{zheng2024fdvision}\newline $\bullet$ Monte Carlo-based uncertainty estimation handles variability effectively~\cite{huang2024mambamir} \\ \hline
    
    \rowcolor{magenta!10} \textbf{Medical Language Processing} & ClinicalMamba~\cite{yang2024clinicalmamba} & Leveraging a pre-trained Mamba language model, precisely analyzes longitudinal clinical notes, capturing temporal relationships and context within patient records. It employs a three-step methodology & $\bullet$ Enhances context understanding for better medical insights by utilizing prompt-based fine-tuning~\cite{yang2024clinicalmamba}\newline  $\bullet$ Facilitates accurate cohort selection in studies and efficiently handles ICD coding tasks~\cite{yang2024clinicalmamba} \\ \hline
    
    \rowcolor{lime!10} \textbf{Medial Image Segmentation - Pure} & Mamba-UNet~\cite{wang2024mambaunet} \\ TM-UNet~\cite{tang2024rotate} \\ Mamba-HUNet~\cite{sanjid2024integrating} \\ VM-UNet~\cite{ruan2024vmunet} \\ Vivim (Video)~\cite{yang2024vivim} \\ ViM-UNet~\cite{archit2024vimunet} & Excel at capturing long-range dependencies and intricate spatial relationships within medical images by utilizing visual SSM modules & $\bullet$ Discerning long-range dependencies and modeling complex spatial relations for accurate segmentation~\cite{wang2024mambaunet, tang2024rotate, ruan2024vmunet,archit2024vimunet}\newline $\bullet$ Achieves efficient segmentation with linear time complexity~\cite{wang2024mambaunet, yang2024vivim, sanjid2024integrating}\\ \hline
    
    \rowcolor{teal!10} \textbf{Medial Image Segmentation - Hybrid} & U-Mamba~\cite{ma2024umamba} \\ UltraLight VM-UNet~\cite{wu2024ultralight} \\ ProMamba~\cite{xie2024promamba} \\ H-vmunet~\cite{wu2024hvmunet} \\ VM-UNET-V2~\cite{zhang2024vmunetv2} \\ LMa-UNet~\cite{wang2024large} \\ LightM-UNet~\cite{liao2024lightmunet} \\ P-Mamba~\cite{ye2024pmamba} \\ Swin-UMamba~\cite{liu2024swinumamba} \\ SegMamba~\cite{xing2024segmamba} \\ T-Mamba~\cite{hao2024tmamba} \\ HC-Mamba~\cite{xu2024hcmamba} & Hybrid models are proposed to combine the strengths of convolutional blocks and SSMs, enhancing local representation and feature capturing. & $\bullet$ Balances local and global representation effectively by introducing convolutional blocks~\cite{ma2024umamba,liu2024swinumamba, zhang2024vmunetv2, ye2024pmamba}\newline $\bullet$ Integrating low and high-level features seamlessly~\cite{hao2024tmamba, ye2024pmamba,zhang2024vmunetv2}\newline $\bullet$ Reducing computational complexity for faster processing~\cite{wu2024ultralight, liao2024lightmunet, xing2024segmamba, zhang2024vmunetv2}\\ \hline
    
    \rowcolor{blue!10} \textbf{Medial Image Segmentation - Other Learning Types} \\ \textcolor{gray}{\textit{Weakly-Supervised Semi-Supervised \\ Self-Supervised}} & Weak-Mamba-UNet~\cite{wang2024weakmambaunet} \\ Semi-Mamba-UNet~\cite{ma2024semimambaunet} \\ ViM4Path~\cite{nasirisarvi2024vim4path} & Alternative learning methods, including self-supervised, semi-supervised, and weakly supervised techniques, aim to improve segmentation performance with limited supervision beside utilizing visual SSM modules. & $\bullet$ Effectively handles segmentation tasks with limited annotated data~\cite{wang2024weakmambaunet,ma2024semimambaunet,nasirisarvi2024vim4path} \newline $\bullet$ Leveraging self-supervised and semi-supervised learning to improve segmentation performance~\cite{nasirisarvi2024vim4path,ma2024semimambaunet}\newline $\bullet$ Applying  weakly supervised learning for better feature utilization and improving segmentation accuracy~\cite{wang2024weakmambaunet}\\ \hline
    \end{tabular}
}

\label{table:1}
\end{table}

\Cref{table:1} highlights the key concepts and objectives of the reviewed Mamba-based models according to each category, representing the core ideas and important features that can be investigated and utilized in future research.

Mamba-based models demonstrate significant advancements across various medical imaging tasks by leveraging shared representations and advanced techniques. In multi-task learning, models like MambaMIL \cite{yang2024mambamil} and nnMamba \cite{gong2024nnmamba} enhance generalization and reduce overfitting, proving effective in segmentation, classification, and landmark detection. Multi-modal models such as SurvMamba \cite{chen2024survmamba} and MambaDFuse \cite{li2024mambadfuse} integrate pathological images with genomic data, improving survival predictions and clinical outcomes through high-quality image fusion. Medical image Synthesis models, including MD-Dose \cite{fu2024mddose} and VM-DDPM \cite{ju2024vmddpm}, combine local and global modelling to manage noise and extract structural information efficiently, enhancing dose prediction and image synthesis. Registration models like VMambaMorph \cite{wang2024vmambamorph} and MambaMorph \cite{guo2024mambamorph} utilize high-dimensional feature learning for accurate and fast alignment of medical images. Reconstruction models, such as FD-Vision Mamba \cite{zheng2024fdvision} and MambaMIR \cite{huang2024mambamir}, employ dual-path designs to enhance image quality and handle variability. ClinicalMamba \cite{yang2024clinicalmamba} uses a pre-trained language model in medical language processing to analyze clinical notes, capturing temporal relationships and context, thereby improving cohort selection and ICD coding tasks. These models collectively contribute to improved accuracy, efficiency, and clinical insights in medical image analysis.

Much of the research has focused on medical image segmentation (particularly in proposing hybrid networks), where inputs often entail high-resolution data. These segmentation frameworks leverage novel convolutional blocks to efficiently capture local representations while utilizing SSM blocks to capture long-range dependencies seamlessly. For example, U-Mamba \cite{ma2024umamba} employs Mamba blocks followed by residual convolutional blocks to understand local and global information sequentially. Similarly, for segmenting 2D and 3D medical images, LMa-UNet \cite{wang2024large} uses large windows, transforming local spatial modelling compared to traditional methods. This method is remarkably efficient in global modelling, surpassing self-attention but with linear computational complexity. It introduces a novel hierarchical and bidirectional Mamba block that enhances global modelling capabilities and proficiency in capturing intricate spatial relationships while using a convolutional decoder to reconstruct the exact spatial relationships. Two approaches have been proposed to reduce computational complexity significantly \cite{wu2024ultralight,liao2024lightmunet}. Notably, UltraLight VM-UNet \cite{wu2024ultralight}, another hybrid approach, presents a lightweight parallel SSM module alongside convolutional blocks to enhance local and global segmentation accuracy. It is worth noting that pure models attempt to leverage different visual Mamba blocks as an alternative to transformer blocks and self-attention mechanisms, which suffer from the quadratic computational burden. In both segmentation categories, SSM blocks capture long-range dependencies across various datasets in different modalities. However, they have not been effectively modified to act as versatile modules capturing both types of information efficiently or to serve as a trade-off. In recent developments, the "Other Learning Approaches" category has tried to apply Mamba-based networks with novel learning types to reduce the need for vast amounts of labelled training data and maximize feature extraction from labelled and unlabeled data.

Like medical image segmentation, Mamba is also used in medical image classification across different modalities to capture global dependencies and utilize novel learning methodologies. 
Classification models like CMViM \cite{yang2024cmvim}, and MedMamba \cite{yue2024medmamba} improve accuracy by integrating convolutional layers with SSMs. Techniques such as contrastive learning and multiple instance learning (MIL) enhance the robustness of 3D medical data classification. These models are designed for efficient GPU usage, making them suitable for processing large-scale medical datasets while maintaining high accuracy. 

Despite the advancements in various Mamba-based models, there remains a need to investigate why these models have gained popularity in medical imaging and why some tasks have more successfully adopted them. SSM-based models, particularly in tasks like segmentation, have become increasingly favoured due to significantly reducing the model's theoretical complexity and its modelling capability on high-resolution data. In medical imaging, efficiently processing high-resolution images that provide accurate local information for disease detection is crucial, and these models excel in this area. While SSMs have the potential to be applied to different tasks, they may require further modifications to be
adapted to other specific tasks. For example, while being applied to classification, the main contributions in this task still lie under the broader ideas of leveraging these methodologies rather than modifying the core SSM structures to achieve SOTA results \cite{yu2024mambaout}. Based on \Cref{fig:taxonomy}, there is still room for studies concerning other medical tasks in particular categories, such as anomaly detection and image denoising. Lastly, it is worth underscoring the necessity for trailblazing efforts in this dynamic and swiftly advancing field, with numerous papers currently emerging alongside our survey \cite{atli2024i2i,fallahpour2024ehrmamba,xu2024hc,nguyen2024ac,sanjid2024optimizing}. We are committed to continuously updating our survey to reflect these ongoing developments.

\section{Future Direction and Open Challenges}
\label{sec:future-direction}

Throughout this survey, we have thoroughly analysed several Mamba models, exploring their architectural designs, motivations, objectives, and applications, all focused on addressing real-world challenges. In this section, we delve into the challenges linked to Mamba's unexplored potential in the medical domain. Our goal is to pinpoint critical areas where Mamba's capabilities can be enhanced and expanded, thereby contributing to the continuous advancement of medical Mamba.

$\bullet$~ \textbf{\textit{Generalization:}} 

The power of generalization is key in developing smart healthcare, enabling a wide range of research and clinical diagnosis applications. However, the Mamba structure can be improved to enhance its ability to generalize. By effectively managing domain-specific information that may accumulate or even amplify within hidden states, we can significantly improve the generalization performance of the SSM model in medical image analysis.

$\bullet$~ \textbf{\textit{Scanning Mechanism:}} 


The selective scanning mechanism is a fundamental aspect of the Mamba model, originally optimized for 1D causal sequential data. While effective for autoregressive tasks, this mechanism is less suitable for understanding visual domains~\cite{yu2024mambaout}. Many existing approaches utilize bi-directional scanning and extend scanning directions to capture the spatial information inherent in 2D or higher-dimensional medical visual data to address the non-causal nature of visual data. Despite these adaptations, Mamba-based models have shown a performance drop in mainstream CV tasks and face a performance gap compared to SOTA CNN-based and ViT models in high-level vision tasks. Therefore, there is a potential need for more innovative scanning schemes to fully harness the potential of higher-dimensional non-causal visual data, especially in medical imaging applications.

$\bullet$~ \textbf{\textit{Explainability:}} 

Studies have experimentally elucidated the mechanisms underlying the Mamba model and its SSM counterparts in NLP, focusing on their in-context learning capabilities~\cite{grazzi2024mamba}. Additionally, other works have established theoretical foundations for Mamba's applications in NLP~\cite{cirone2024theoretical}, analyzing the Mamba model and other pioneering SSMs from a control theoretic perspective to unravel the decision-making processes within these models~\cite{alonso2024state}. Despite these advancements, understanding why Mamba excels in visual tasks, particularly in medical imaging, remains challenging. This underscores the need for further research to enhance the interpretability of Mamba networks in the field of medical image analysis, aiming to unravel the nuances of exploitability in visual Mamba networks.

$\bullet$~ \textbf{\textit{Inferior Performace on Some Tasks:}}

Recent studies suggest that SSMs are potentially beneficial for tasks that follow the long-sequence and autoregressive characteristics such as object detection \& segmentation \cite{yu2024mambaout}, while performing inferior in image classification, which does not benefit from either feature. Given the significance of models pre-trained on extensive datasets like ImageNet \cite{deng2009imagenet} for various downstream applications, releasing the base, large, and huge versions of Mamba could be extremely beneficial and represent a promising future direction. 
Recently, the authors of Mamba have introduced the Mamba-2 architecture \cite{dao2024transformers}, which combines SSMs with attention mechanisms. This new design leverages a theoretical framework that effectively bridges the gap between SSMs and attention, significantly enhancing Mamba’s performance.

$\bullet$~ \textbf{\textit{Adaptation of Mamba to Medical Foundation Models:}}

Large-scale pre-trained models, namely foundation models \cite{azad2023foundational}, are crucial in the field of medical imaging and considered an important step towards artificial general intelligence (AGI) \cite{awais2023foundational}. Although there are efforts to incorporate Mamba as a foundation model, such as the Jamba framework \cite{lieber2024jamba} for language modelling, there is still significant potential for developing foundational models based on Mamba specifically for medical imaging. This area remains largely unexplored, presenting a vast opportunity for innovation and improvement in medical imaging analysis.

$\bullet$~ \textbf{\textit{Multi-modal Mamba:}}

Medicine is inherently a multi-modal field containing various sources of information, such as clinical notes, laboratory tests, medical images, and more, to deliver patient care \cite{tu2024towards}. Despite significant advancements in biomedical AI, most existing models, including those based on mamba, are still unimodal and focused on single tasks. Consequently, there is a need to investigate further the development of new SSM-based backbones for the multi-modal learning paradigm.

\section{Conclusion}
\label{sec:conclusion}

This survey reviews state space models (SSMs) in the medical domain, focusing on the \textbf{\textit{Mamba}} architecture. Specifically, the applications of Mamba models in image segmentation, reconstruction, registration, classification, language processing, multi-modal understanding, and other medically related works were taxonomized and fully explored from multiple perspectives. We provide systematic categorization of the literature incorporating additional facets such as imaging modalities, organs of interest, and the computational complexity of these methods. Furthermore, we consolidate the potential future direction and the existing challenges, which should be carefully
considered in practice, can inspire more innovative works
within the field of medical image analysis. It is worth mentioning that some of the papers referenced in this survey are pre-prints, particularly relevant because much of the work in this field is happening simultaneously. Nevertheless, we have diligently ensured the inclusion of high-quality research from respected sources in this fast-moving area. We aspire this endeavour to offer a roadmap for researchers in medical imaging, harnessing the potential of Mamba while also addressing its shortcomings.


\bibliographystyle{unsrt}  
\bibliography{references}

\end{document}